\documentclass[sigconf, nonacm]{acmart}

\newcommand\vldbdoi{XX.XX/XXX.XX}

\newcommand\vldbvolume{14}
\newcommand\vldbissue{1}

\newcommand\vldbavailabilityurl{https://github.com/BU-DiSC/osmtree}
\newcommand\vldbpagestyle{plain} 

\usepackage{algorithm2e}
\usepackage{amsmath}
\usepackage{multicol}
\usepackage{hyperref}

\setcopyright{rightsretained}

\acmDOI{}

\acmISBN{}

\acmConference{Under Submission} 
\acmYear{2022}
\copyrightyear{}
\acmPrice{}
\acmSubmissionID{}
\usepackage{balance}
\usepackage{color}
\usepackage{graphicx} 
\usepackage{amsmath} 
\usepackage{algorithm2e} 
\usepackage[noend]{algorithmic} 
\usepackage{enumitem}
\usepackage{booktabs, multirow}

\algsetup{linenosize=\scriptsize}

\usepackage{fixme}
	\fxsetup{
    	status=final,
    	author={Note},
	    layout=pdfnote,
    	theme=color,
	    silent=true,
    	inline=false
	}

\newcommand{\showDOI}[1]{\unskip}
\newcommand{\showURL}[1]{\unskip}

\usepackage{wrapfig}
\usepackage{subcaption}

\usepackage{moresize}
\usepackage{cleveref}
\crefname{section}{\S}{\S}
\Crefname{section}{\S}{\S}

\definecolor{myRed}{rgb}{0.73, 0.31, 0.28}
\definecolor{myBlue}{rgb}{0, 0.44, 1}
\definecolor{myGreen}{rgb}{0.31, 0.78, 0.47}
\definecolor{myGrey}{rgb}{0.57, 0.64, 0.69}

\newcommand{\Paragraph}[1]{\vspace{0.8 mm} \noindent \textbf{#1.}}
\newcommand{\Paragraphit}[1]{\vspace{0.8 mm} \noindent {\emph{\underline{\smash{#1.}}}}}

\newcommand{\Paragraphnopunc}[1]{\vspace{1 mm} \noindent {\bf #1}}

\usepackage{mathtools}

\newcommand{\eat}[1]{}

\usepackage{fixme}
\fxsetup{
    status=draft,
    author={Note},
    layout=inline,
    theme=color,
    silent=true,
    inline=true
}
\FXRegisterAuthor{ma}{ama}{\color{red}Manos}
\FXRegisterAuthor{kb}{kba}{\color{blue}Kenneth}
\FXRegisterAuthor{si}{sia}{\color{green}Stratos}
\FXRegisterAuthor{tk}{tka}{\color{red}to Kenneth}
\FXRegisterAuthor{ts}{tsa}{\color{magenta}to Stratos}
\FXRegisterAuthor{tsw}{tswa}{\color{red}to Shijia}

\usepackage{smartdiagram}
\usepackage{tikz}
\usetikzlibrary{arrows,positioning}

\usepackage[utf8]{inputenc}
\usepackage[ampersand]{easylist}
\usepackage{mathtools}
\usepackage{color}
\usepackage{listings}
\usepackage{caption}
\usepackage{hyperref}

\newcounter{nalg}[section] 
\renewcommand{\thenalg}{\thechapter \arabic{nalg}} 
\DeclareCaptionLabelFormat{algocaption}{Algorithm \thenalg} 

\newcommand{\sysName}[0]{\textcolor{black}{\sysNameTitle{}}}
\newcommand{\sysNameTitle}[0]{OSM-tree}
\newcommand{\pinBufName}[0]{\textcolor{black}{OSM-buffer}}
\newcommand{\bplustree}[0]{B$^+$-tree}

\newcommand{\bepsilontree}[0]{\textcolor{black}{B$^\epsilon$-tree}}

\newcommand{\lastsortedzone}[0]{\textcolor{black}{\texttt{last\_sorted\_zone}}}

\newcommand{\previousboundary}[0]{\textcolor{black}{\texttt{previous\_boundary}}}

\begin{document}

\title{\sysNameTitle{}: A Sortedness-Aware Index}




\author{Aneesh Raman$^1$, Subhadeep Sarkar$^1$, Matthaios Olma$^2$, Manos Athanassoulis$^1$}
\affiliation{
   \vspace*{1mm}
   \institution{$^1$Boston University, $^2$Microsoft Research}
   \country{}
}

\begin{abstract}
Indexes facilitate efficient querying when the selection predicate is on 
an indexed key. As a result, when loading data, if we anticipate future
selective (point or range) queries, we typically maintain 
an index that is gradually populated as new 
data is ingested. In that respect, \textit{indexing
can be perceived as the process of adding 
structure} to an incoming, otherwise unsorted, data 
collection. The process of adding structure comes at a cost, as instead of 
simply appending incoming data, every new entry
is inserted into the index. If
the data ingestion order matches the indexed 
attribute order, the ingestion cost is entirely 
redundant and can be avoided (e.g., via bulk loading in a \bplustree). However, state-of-the-art index
designs do not benefit
when data is ingested in an order that is 
\textit{close to} being sorted but \textit{not} fully sorted. 

In this paper, we study how indexes 
can benefit from \textit{partial data sortedness} or \emph{near-sortedness}, and we propose
an ensemble of techniques that combine \textit{bulk loading}, \textit{index appends}, \textit{variable node fill/split 
factor}, and \textit{buffering}, to optimize the ingestion
cost of a tree index in presence of partial data sortedness. 
We further augment the proposed design with necessary 
metadata structures to ensure competitive read performance. 
We apply the proposed design paradigm on a 
state-of-the-art 
\bplustree, and we propose the
Ordered Sort-Merge tree (\sysName{}). \sysName{} outperforms 
the state of the art by up to $8.8\times$ in ingestion performance in the
presence of sortedness, while falling back to a \bplustree{}'s
ingestion performance when data is scrambled. \sysName{}
offers competitive query performance, leading to performance benefits 
between $28\%$ and $5\times$ for mixed read/write workloads.

\end{abstract}

\maketitle

\pagestyle{\vldbpagestyle}
\vspace{-0.3in}
\begingroup
\renewcommand\thefootnote{}\footnote{\noindent
This work is licensed under the Creative Commons BY-NC-ND 4.0 International License. Visit \url{https://creativecommons.org/licenses/by-nc-nd/4.0/} to view a copy of this license. For any use beyond those covered by this license, obtain permission by emailing \href{mailto:info@vldb.org}{info@vldb.org}. Copyright is held by the owner/author(s). Publication rights licensed to the VLDB Endowment. \\
\raggedright Proceedings of the VLDB Endowment, Vol. \vldbvolume, No. \vldbissue\ %
ISSN 2150-8097. \\
\href{https://doi.org/\vldbdoi}{doi:\vldbdoi} \\
}\addtocounter{footnote}{-1}\endgroup

\ifdefempty{\vldbavailabilityurl}{}{
\vspace{.3cm}
\begingroup\small\noindent\raggedright\textbf{PVLDB Artifact Availability:}\\
The artifacts have been made available at \url{\vldbavailabilityurl} and \url{https://github.com/BU-DiSC/sortedness-workload}.
\endgroup
}

\vspace{-0.08in}
\section{Introduction}
\label{sec:introduction}
\vspace{-0.02in}

Database indexing sits at the heart of almost any data system varying from
full-blown relational systems \cite{Ramakrishnan2002} to NoSQL key-value 
stores \cite{Idreos2020}. Indexes help accelerate query processing both 
for analytical and transactional workloads by allowing efficient data
accesses of selective (range or point) queries. 
Essentially, in presence
of read queries, database administrators decide to build and maintain 
indexes to improve query performance at the expense of space and write 
amplification \cite{Athanassoulis2016}, and the 
time needed to update the indexes.


\begin{figure}[t]
	\centering
	\includegraphics[width=\columnwidth]{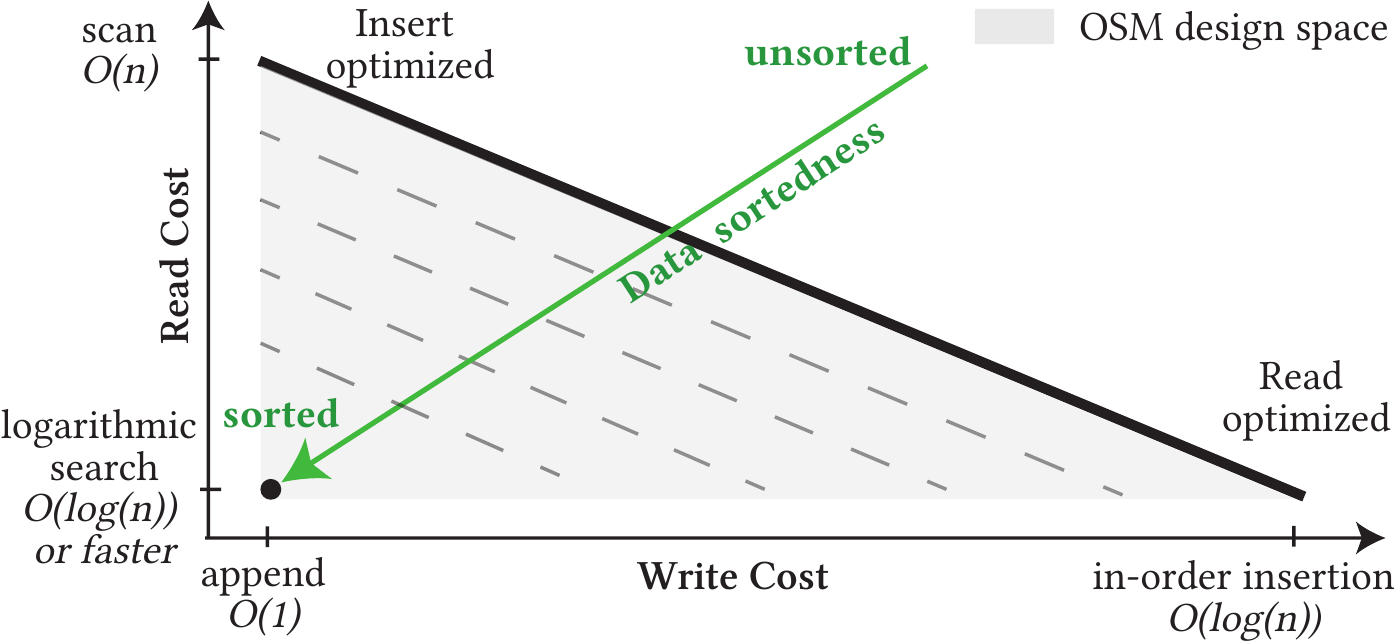}
	\caption{State-of-the-art indexing and data organization techniques pay a higher write cost in order to store data as sorted (or, in general, more organized) and offer efficient reads. Since the goal of indexing is to store the data as sorted, we ideally expect that ingesting \emph{near-sorted} data would be more efficient, which is not the case. We introduce the OSM meta-design that offers better performance as data exhibit higher degree of sortedness.}
	\label{fig:w-r-intro}
	\vspace{-0.2in}
\end{figure}

\Paragraph{Indexing Adds Structure to Facilitate Queries}
We pay the cost of index construction and maintenance 
because it \emph{adds structure} to the data, which, in turn, allows for
efficient queries. As shown in Figure \ref{fig:w-r-intro}, with the thick 
black
line, every index or, in general, any data organization technique, exhibits a 
fundamental tradeoff between its \textit{read} and \textit{write} cost. 
To achieve efficient logarithmic search time for point
queries, a classical index would insert data in their correct position 
(\emph{in-order insertion}) leading to efficient \emph{logarithmic search} (bottom right part of the figure). On the other extreme, if read 
queries are infrequent, then, scanning is acceptable and instead of adding
new entries to an index, we can simply append them
(\emph{appends} leading to \emph{scans} in the top left part of the figure).
This read vs. write tradeoff holds for any data organization effort, including,
classical indexing like \bplustree{} \cite{Graefe2011}, write-optimized 
log-structured merge (LSM) trees \cite{Luo2020b}, or simple online sorting via 
in-order insertion. Since all such data
organization efforts are essentially \emph{adding structure to an otherwise
unstructured data collection}, one would expect they benefit when such
structure exists already. Indeed, if the data entries are already fully
sorted, then we can benefit, for example, by bulk loading a \bplustree. 
However, a question that remains open is \emph{what happens when the data 
entries arrive with some degree of structure but are not entirely sorted} (e.g., 
due to implicit 
clustering~\cite{Athanassoulis2014,Moerkotte1998}). Before
addressing this question, we first explain the concept of \emph{data 
sortedness} and showcase real-life examples of
workloads with variable degree of data sortedness.

\begin{figure}[t]
     \centering
     \begin{subfigure}[b]{0.5\columnwidth}
         \centering
         \includegraphics[height=1.2in]{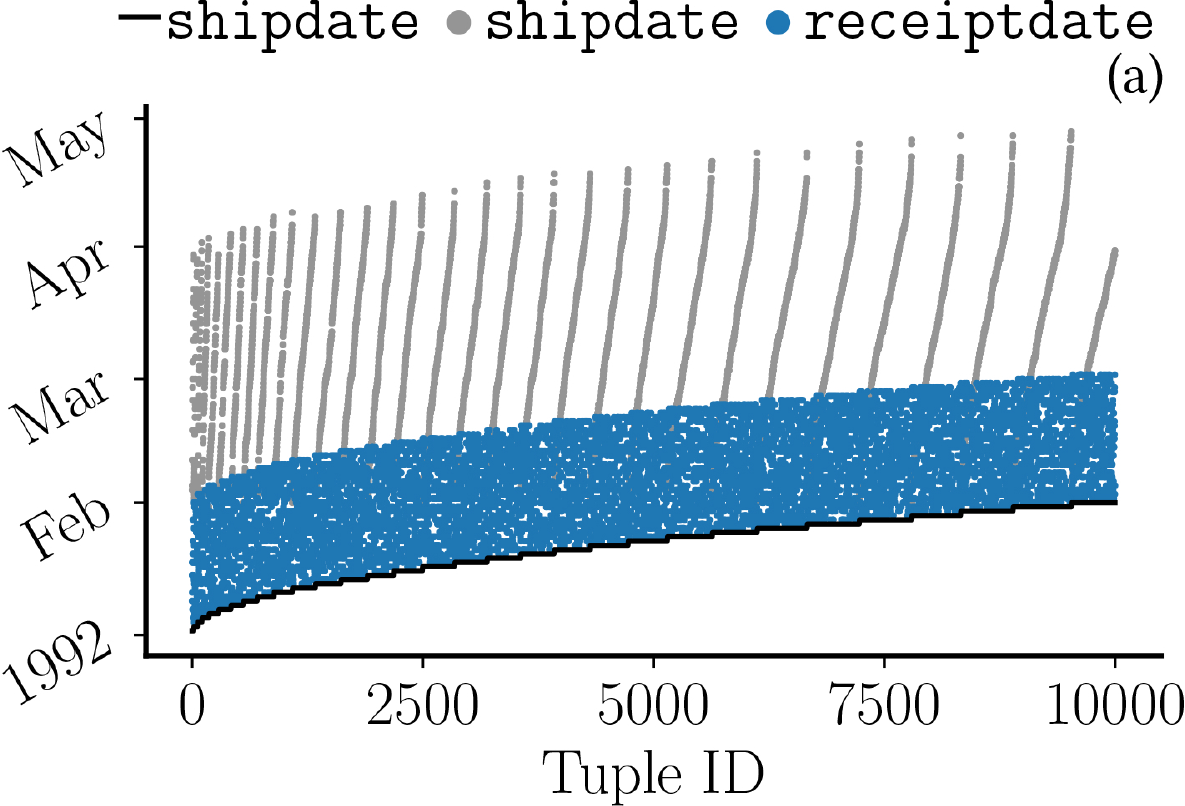}
     \end{subfigure}
     ~
     \begin{subfigure}[b]{0.5\columnwidth}
         \centering
         \vspace{-2mm}
         \includegraphics[height=1.2in]{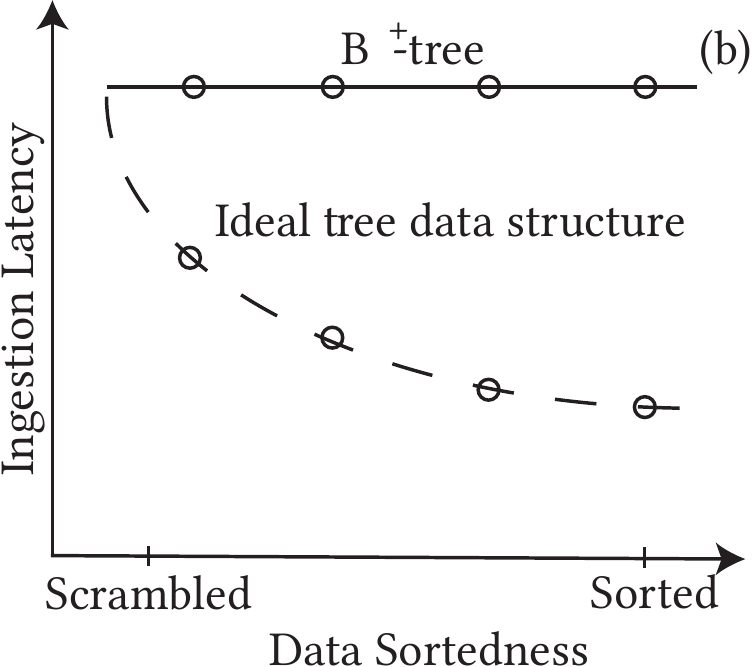}
     \end{subfigure}
        \caption{(a) TPC-H implicit clustering between \texttt{shipdate}, \texttt{commitdate}, and \texttt{receiptdate} leads to near-sorted columns when the data is sorted based on one of them. (b) Ideally, index insertion performance should improve when inserting already sorted or near-sorted data.}
        \label{fig:intro}
        \vspace{-0.1in}
\end{figure}

\Paragraph{Data Sortedness}
Data Sortedness can be defined as the \emph{arrival order} of the indexed key 
upon ingestion. This arrival order can be fully sorted, 
nearly sorted, less sorted, or scrambled. 
Data entries may be \emph{near-sorted} in several real-world cases. Consider the 
TPC-H~\cite{TPCH} \texttt{lineitem} table that has three date-related attributes. 
Figure~\ref{fig:intro}(a), which depicts the first $10,000$ values of 
\texttt{shipdate}, \texttt{commitdate}, and \texttt{receiptdate} of the 
lineitem table, shows that when the data 
arrives in order based on \texttt{shipdate}, the other two attributes are very close to also 
being sorted. There are several scenarios that lead to near-sorted data collections. For
example, (i) a relation that was sorted but a few new arbitrary updates
took place, (ii) data that has been created based on a previous operation or
query, e.g., a join, (iii) data that is sorted based on another attribute 
that is naturally correlated (like the TPC-H example above), or (iv) the 
timestamp attribute of an incoming data stream that has a few data packets 
arriving out of order due to network congestion \cite{Ben-Moshe2011}. 

In order to be
able to exploit this near-sortedness, we need a way to quantify it. In fact,
there have been multiple sortedness metrics 
proposed~\cite{Ben-Moshe2011,Mannila1985,Carlsson1992}. Several of the 
sortedness metrics focus on quantifying the number of inversions or
transpositions needed to achieve full 
sortedness~\cite{Ajtai2002,Cormode2001,Gopalan2007,Gupta2003,Mannila1985}. 
However, a more natural
way of thinking about the \emph{degree of sortedness} in the context of indexing is \emph{how many elements}
are in the ``wrong'' position and, more crucially, \emph{by how much}. As a result,
we use the $(K,L)$ sortedness metric \cite{Ben-Moshe2011} that quantifies
sortedness using two parameters: $K$, that captures the number of elements that
are out of order and $L$, that captures the maximum displacement in terms
of position of the out of order elements. Going back to the TPC-H
example in Figure~\ref{fig:intro}(a), when the data is sorted on
\texttt{shipdate}, \texttt{commitdate} has {$K=99.2\%$} and 
{$L=1.6\%$}, while \texttt{receiptdate} has higher 
sortedness with {$K=96.7\%$} and {$L=0.1\%$}. 
For the latter, this means that 96.7\% of the entries are not at the position they would be if they were sorted and the maximum displacement 
is 0.1\% of the data size.
The $(K,L)$ metric helps quantify the 
number of entries we need to 
work on to \emph{absorb} sortedness
when inserting new data.

\Paragraph{Problem: Indexes Do Not Exploit Data Near-Sortedness}
While indexes can already benefit from inserting a fully sorted data collection
via bulk loading~\cite{Achakeev2013,DenBercken2001} (assuming that it is already known that data is sorted), they
are not designed to exploit near-sortedness. Further, widespread buffering
techniques to optimize index insertion (like in 
\bepsilontree{s}~\cite{Brodal2003} and LSM-trees~\cite{ONeil1996}) are not 
designed to exploit sortedness and will end up sorting the data even when they
are (almost) sorted. We argue that when inserting data to an index, the 
\emph{higher the data sortedness, the lower the insertion cost should be},
as depicted in Figure~\ref{fig:intro}(b), for the \emph{ideal tree data 
structure}. 
Note that state-of-the-art indexes, like \bplustree{s}, do not exhibit any
performance improvement when inserting near-sorted data. In fact, if data is 
inserted in order and bulk loading is not employed, a \bplustree{} would have
the worst-possible space amplification, since every node will be exactly 50\% 
full. In contrast, a \emph{sortedness-aware index} should achieve a better read vs. write tradeoff as \emph{data sortedness 
increases}, as depicted
by the dashed lines in Figure~\ref{fig:w-r-intro}. Following the green line,
which constitutes a third axis -- the one for sortedness -- we envision a new 
class of data structures that is able to navigate the entire shaded 
region in Figure~\ref{fig:w-r-intro} and perform ``less'' indexing for 
near-sorted data, leading up to the ideal performance of 
append-like cost of insertion with efficient searching if data
is sorted.


\Paragraph{Our Approach: \sysName{}}
To realize this vision, we propose a new paradigm for designing indexes that is capable of exploiting any existing degree of
sortednes to improve index insertion performance without hurting read latency. 
We achieve this using an ensemble of techniques which, when combined appropriately, solve
a problem that cannot be solved by any one of them alone. Specifically, we employ 
\emph{buffering}, \emph{partial bulk loading}, \emph{query-driven partial sorting}, and
\emph{merging} to create an index ingestion mechanism that substantially 
accelerates the ingestion performance in the presence of data sortedness. This sortedness-aware 
approach, however, comes at the cost of increased read latency since every query may have to
search a buffer. To alleviate this cost, we augment the design with a collection of 
Zonemaps~\cite{Moerkotte1998} and Bloom filters (BFs)~\cite{Bloom1970},
which help make the read cost comparable to the baseline. By combining the ingestion and
the read optimizations on top of a state-of-the-art \bplustree{}, we propose our new
design termed Ordered Sort-Merge tree (\emph{\sysName{}}). In a nutshell, \sysName{}
buffers incoming data to bulk load as much as possible and reverts back to insertion
from the root (\emph{top-inserts}) otherwise. By adaptively sorting buffered data during queries, \sysName{} 
avoids the burden of sorting large data collections. With respect to reads, it uses
interpolation search for the sorted parts of the buffer, and pays the cost of scanning
a small amount of data when the Zonemaps and Bloom filters direct a query to 
unsorted entries in the buffer. Note that 
\textbf{\emph{the OSM
paradigm can make any tree-based data
structure (e.g., radix trees, \bepsilontree{s}, and LSM-trees) amenable to 
data sortedness}}. It is not a new index \emph{per se}, rather, a new framework for creating sortedness-aware counterparts for any tree-based index.


\Paragraph{Contributions} Our work offers the following contributions.
\vspace{-0.03in}
\begin{itemize}[leftmargin=1.3em, itemsep=0.25em]
	\item We identify \emph{sortedness as a resource} that can be harnessed to ingest data faster in tree indexes. 
	\item We propose a new index \emph{meta-design} that employs buffering, partial bulk loading, and merging to enhance ingestion in the presence of any degree of data sortedness. 
	\item We augment this design to propose \sysName{} that encompasses query-driven sorting, merging, Zonemaps, and Bloom filters to maintain competitive performance for point and range queries.
	\item We apply this design on a state-of-the-art \bplustree{}, and we show that we can achieve up to $8.8\times$ faster data ingestion with competitive read query performance leading to performance benefits of up to $5\times$ in mixed read/write workloads.
	\item The OSM meta-design provides the foundation needed to capture a varying degree of data pre-sortedness for tree indexes.
\end{itemize}
\vspace{-0.1in}

\section{Problem Statement}
\label{sec:background}


Both indexing and sorting pay the cost to add
structure to the data to facilitate faster
queries. Both assume that the data is not sorted and that the desired state if fully sorted. \emph{Do we need to pay the same cost (of sorting or indexing) for near-sorted data?}



\begin{center}
\Paragraphnopunc{Challenge:} \emph{Handling Variable Degrees of Sortedness.}
\end{center}
The ingestion complexity for an entry in a \bplustree{} is 
$O(log_F(N))$, \emph{irrespective of the order} of ingested data. 
Although this is beneficial in the worst-case when incoming data 
is completely scrambled, such performance is suboptimal even when 
some amount of data sortedness exists. Not only do \bplustree{s} 
do more work, but also carry their worst space amplification for 
fully sorted inserts. Insertions, in this case, are right-deep, 
while both the internal and leaf nodes are split in half, leaving 
50\% of all index nodes unused. The \bplustree{} and other popular 
modern indexes are not designed to identify data sortedness to do 
less work and improve insert efficiency. 

\begin{center}
\Paragraphnopunc{Goal:} \emph{We set out to design an index 
that offers better ingestion performance for higher 
data sortedness, without hurting 
the performance of read queries.}
\end{center}



\section{Design Elements}
\label{sec:design_elements}

\begin{figure}[t]
        \centering
        \includegraphics[width=\columnwidth]{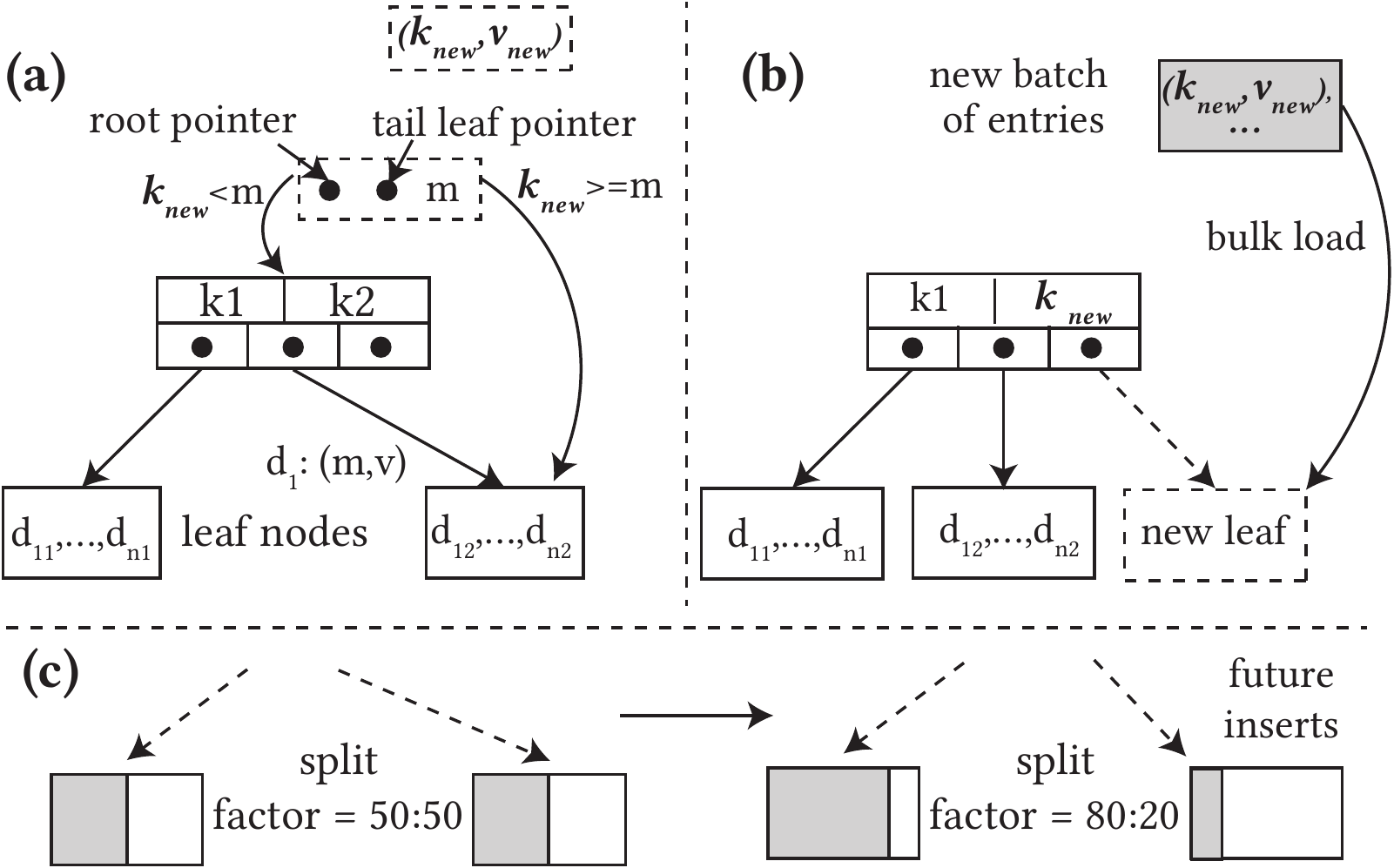}
    \caption{Design elements aimed at exploiting data sortedness.
    (a) An in-order entry can be directly inserted at the tail leaf if $k_{new}$$\geq$$m$.
    (b) If we have a batch of new in-order inserts, they can be bulk loaded to the tree.
    (c) The split factor can be adjusted such that a newly created node from the split reserves 
    more space for following inserts. }
    \label{fig:design-elements}
    \vspace{-0.05in}
\end{figure}

We now present the four fundamental design elements which, when
appropriately combined, allow us to exploit data sortedness. 
The first three: (i) right-most leaf insertions, (ii) bulk loading, 
and (iii) fill/split factor adjustment, benefits as-is a fully
sorted data ingestion, and when combined with (iv) buffering, can lead to a design that can exploit variable data sortedness.
We discuss the key ideas behind the four design
elements and, later in Section \ref{sec:design}, we discuss how
to put them together. We illustrate these ideas in Figure ~\ref{fig:design-elements}.

\Paragraph{Right-Most Leaf Insertion}
When inserting data that follow the order of the index-attribute, 
we can avoid the logarithmic tree traversal cost by always maintaining a pointer to the right-most leaf node, as shown in 
Figure ~\ref{fig:design-elements}(a). That way, for every
new insert, we first check that indeed it should be directed to
the right-most leaf (that is, that the inserted key is larger than
the \emph{minimum} value of that leaf), and we can simply insert 
it in the leaf. Note that this approach can also absorb a very
small degree of sortedness if the size of a leaf node is 
large enough.
In terms of insertion performance, in-order insertion allows us
to have a $O(1)$ insertion cost instead of $O(log_F(N))$. 
In-order insertion can fall-back to classical insert from the
root, which we term \emph{top-inserts}, when the leaf node is
not enough to capture the sortedness. 

\Paragraph{Bulk Loading}
If the data is \emph{fully sorted}, we can perform better than
in-order insertion, by bulk loading the data \cite{DenBercken2001}, 
as shown in Figure ~\ref{fig:design-elements}(b). 
That way, we can avoid accessing a node
for every entry. Instead, we append to an in-memory buffer as
data arrive, and once a page is full, we create a new right-most 
leaf. This amortizes the insertion cost across $F$ entries. While
bulk loading gives great index creation time if data is fully 
sorted, it cannot exploit a varying degree of sortedness.

\Paragraph{Fill Factor/Split Factor Adjustment}
When employing any of the above two techniques, we can further
optimize the shape of the tree by carefully deciding how we split
internal nodes and leaf nodes. Specifically, if the data is fully 
sorted, the classical split algorithm will create half-full nodes 
which will never receive any future inserts. Hence, the nodes
always remain half-full leading the worst-case with respect
to the index space amplification and, in some cases, affecting the
index height as well, since the effective fanout will also be
half of the nominal one. Instead, if we anticipate data to arrive
fully sorted (or as near-sorted as we discuss in the next section), 
we can decide to employ a different split factor where, for 
example, 80\% of the entries would stay on the original node
and the newly created one will only hold 20\% of the data in 
anticipation of the new -- higher in terms of value -- keys, as shown in 
Figure ~\ref{fig:design-elements}(c). 
By changing the split factor, we also allow the nodes to have 
a higher fill factor on average. The split factor change reduces
the number of overall splits needed, improving the insertion
performance. The resulting higher fill factor throughout the 
tree reduces the overall number of nodes needed to hold the 
data which, in turn, reduces the memory footprint of the data structure. 


\Paragraph{Buffering}
The above techniques do not offer substantial benefits
if the data is not fully sorted. To utilize these 
techniques for data with a varying degree of sortedness, we need 
to buffer incoming data to propagate to the tree only those
inserts that are in-order. The buffer is periodically sorted
as new entries arrive, to make the first part eligible for bulk loading. 
In the next 
section,
we discuss how we do this without employing expensive
in-memory sorting, and ultimately, without hurting reads
that may have to access the in-memory buffer to find the requested 
values.

\section{OSM-Tree Design}
\label{sec:design}
In this section, we present a new meta-index design that 
can exploit data sortedness to accelerate ingestion.
Section ~\ref{sec:w-opt-buff-tree} presents the preliminary design that
puts together the three fundamental design elements from 
Section~\ref{sec:design_elements}. In 
Section ~\ref{sec:w-r-opt-buf-tree}, we augment this design
to improve its lookup performance, and finally, in 
Section ~\ref{sec:osm}, we put everything together to present
our novel \sysName{} design.

\subsection{Sortedness-Aware Ingestion}
\label{sec:w-opt-buff-tree}
While right-most leaf insertion and bulk loading can help when the
data is fully sorted, any degree of sortedness essentially would
need a staging area. Hence, we employ a dedicated in-memory
insertion buffer which intercepts all index inserts 
to facilitate future bulk ingestion.
Specifically, the buffer allows bulk loading of multiple data
pages into the tree at a constant cost. 
We now describe the 
buffering mechanism in detail. 

\begin{figure}[!t]
	\centering
	\includegraphics[width=\linewidth]{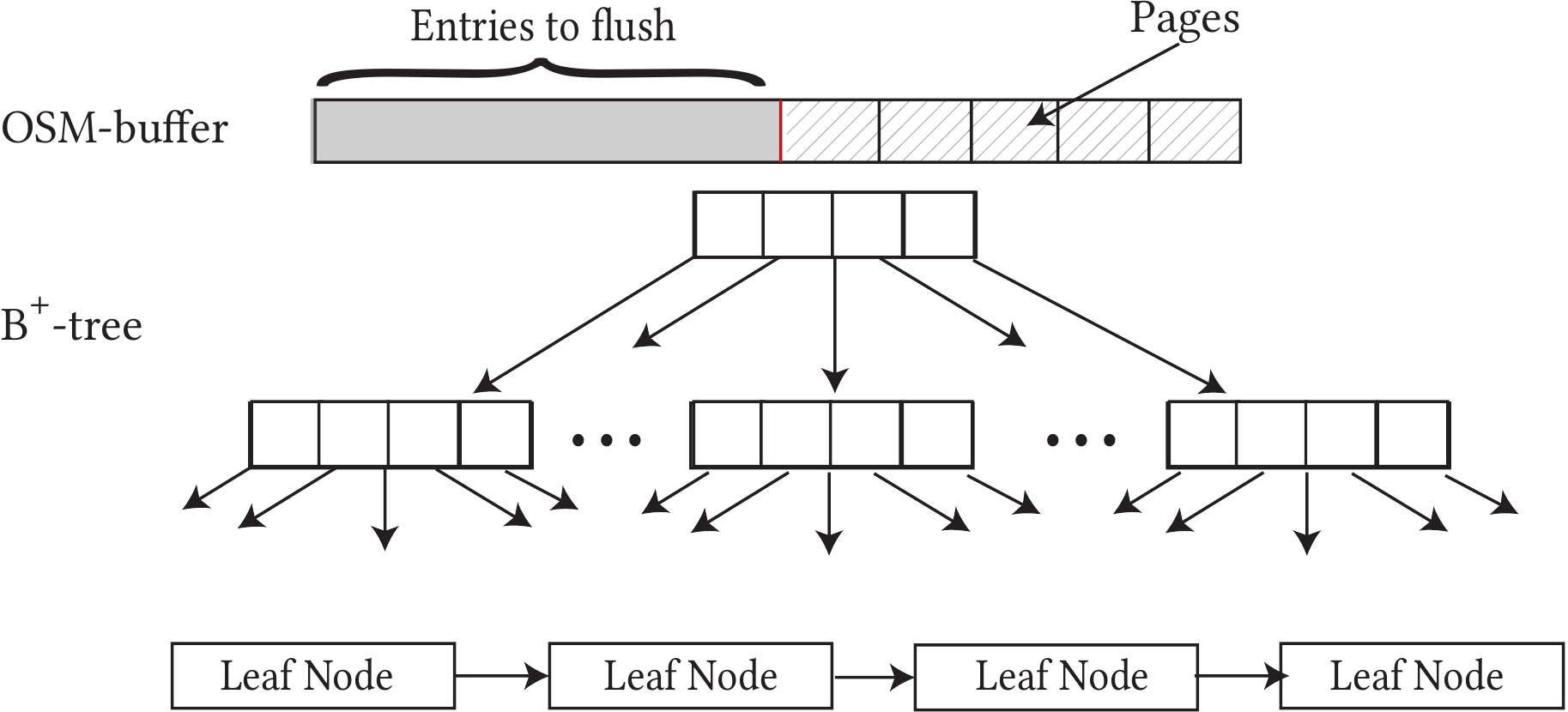}
	
	\caption{The basic design of \sysName{} with a buffer on top of a state-of-the-art \bplustree{}.}
	\label{fig:osm-img}
    \vspace{-0.2in}
 \end{figure}

\Paragraph{Basic Structure}
We assume a state-of-the-art tree index like the \bplustree{}
in Figure ~\ref{fig:osm-img}. Note that any variation of 
\bplustree{} that supports bulk loading can be part of this 
meta-design. On top of the basic index, our design includes a buffer 
that receives incoming data. 

\Paragraph{The OSM-buffer}
The in-memory buffer, termed \textit{\pinBufName{}}, maintains all recently
inserted data and checks whether the entries are inserted in order. In 
general, the data in the buffer is eventually inserted into the index
either through bulk loading, when possible (that is, when the 
buffered data have higher values than the data already in the 
index) or through traditional
inserts from the root node (termed \emph{top-inserts}). By having
this design, we can already guarantee that if data is inserted in 
order, they will be efficiently bulk loaded. 
We now discuss the buffer flushing strategies that optimize data insertion
in the presence of varying degree of sortedness. 

\Paragraph{Flush Strategy}
When the buffer becomes full, we \emph{flush} the buffer to the tree.
This flushing can happen either in the form of \emph{bulk loading}
or in the form of \emph{top-inserts}. Our goal is to maximize
the amount of data inserted into the index via bulk loading.
In the best case -- i.e., when the buffer is fully sorted by virtue
of the pre-existing data sortedness -- we bulk load the contents 
of the buffer with no sorting effort. In general, the buffer may not
be fully sorted, and we would need to sort it before
flushing. At this point, we either (i) bulk load as many pages as possible, if the tree
has strictly smaller values than what the (now sorted) buffer has, 
and (ii) perform top-inserts if there is overlap. Note, that when
we perform top-inserts, after inserting a page worth of data, we re-check for
overlap, and if possible, revert to bulk loading. 

Another decision we make at every flush cycle (i.e., every time the 
buffer is full) is what portion of the buffer to flush. The insight
here is that if we flush the entire buffer, we may insert to the index, entries
that overlap with future inserts if the data is anything but fully sorted. 
Hence, instead of flushing
the entire buffer at every cycle, we flush a portion of the buffer
(by default, half of it). 
Partially retaining entries in the buffer after a flush operation 
helps capture overlaps with future insertions to some extent, which, 
in turn, increases the number of bulk loaded pages (and decreases
top-inserts) across flush cycles.
Note that every top-insert costs $O(log_F(N))$ while bulk loading
costs $O(1/F)$ since $F$ inserts are serviced by a single 
node addition. 

\begin{figure}[t]
    \centering
    \begin{subfigure}[b]{0.23\textwidth}
        \centering
        \includegraphics[scale=0.3]{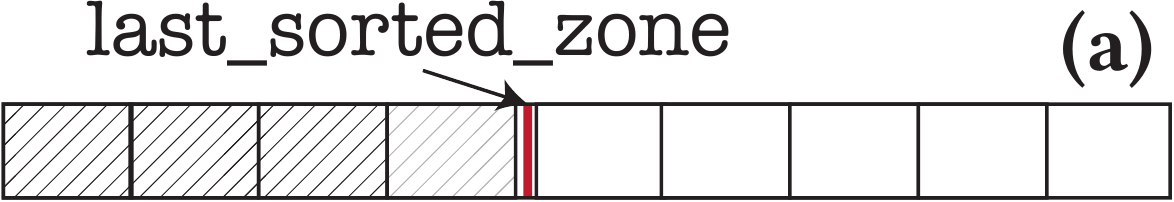}
    \end{subfigure}
    \hfill
    \begin{subfigure}[b]{0.23\textwidth}
        \centering
        \includegraphics[scale=0.3]{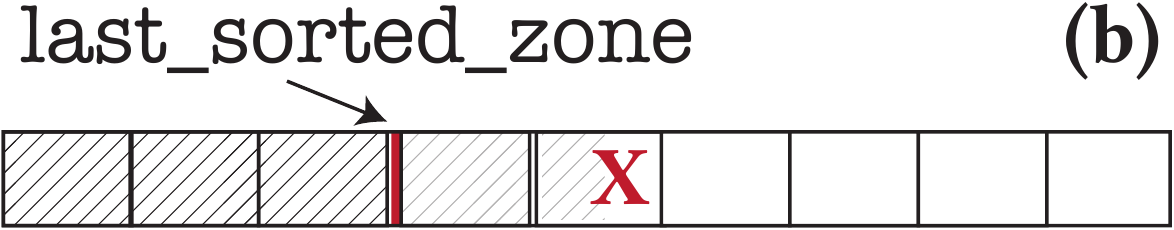}
    \end{subfigure}
    
    \vskip\baselineskip

    \begin{subfigure}[b]{0.23\textwidth}
        \centering
        \includegraphics[scale=0.3]{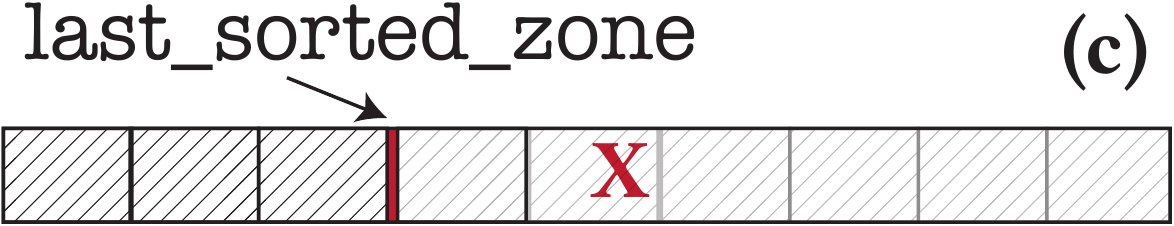}
    \end{subfigure}
    \hfill
    \begin{subfigure}[b]{0.23\textwidth}
        \centering
        \includegraphics[scale=0.3]{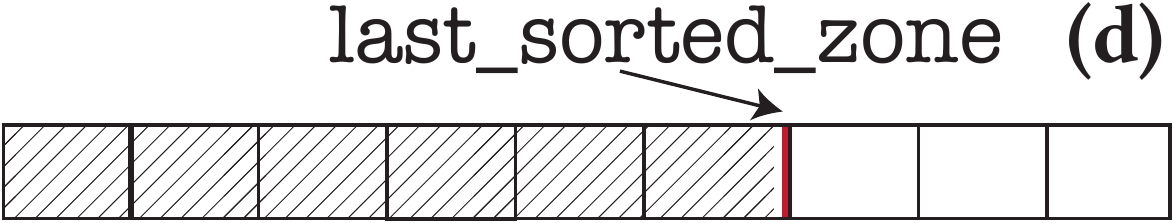}
    \end{subfigure}
\caption{The lifecycle of an insert in \pinBufName{}. 
	(a) \lastsortedzone{} moves as new ordered entries are inserted. 
	(b) An out-of-order entry moves \lastsortedzone{} to the left.
	(c) Newer inserts are added and \lastsortedzone{} only moves if required.
	(d) After a flush, the remaining entries in the buffer are sorted and \lastsortedzone{} 
	is reset to mark the last sorted page in the buffer.}
\label{fig:osmBufInserts}
    \vspace{-0.1in}
\end{figure}

\Paragraph{Zonemaps to Identify Overlaps}
Our design targets ingestion of near-sorted data. In this case,
out-of-order entries are likely to be displaced by a few pages 
from their ideal position. After a flush cycle, the buffer is half
full, and its contents are fully sorted. As a result, no entries are out of order
in the buffer, and we mark the last page containing sorted data
as the \lastsortedzone{} (Figure~\ref{fig:osmBufInserts}(a)). Note that
every buffer page is treated as a separate zone. As new 
entries arrive, because of the potential displacement, a 
newly appended entry may be either (i) overlapping with data in
earlier pages, hence moving the \lastsortedzone{} to the left 
(Figure~\ref{fig:osmBufInserts}(b)), (ii) overlapping without having
to move the \lastsortedzone{} (Figure~\ref{fig:osmBufInserts}(c)), 
or (iii) strictly greater, thus moving the \lastsortedzone{} to the 
right (Figure~\ref{fig:osmBufInserts}(d)). To update
the \lastsortedzone{}, we also maintain Zonemaps per page that
allow for a quick overlap test after every insertion. 
Maintaining the \lastsortedzone{} accurately, helps to avoid unnecessary sorting
at every flush cycle. 

When the buffer becomes full, we use the \lastsortedzone{}
to decide how much to flush. If the most recent entries have moved the 
\lastsortedzone{} to correspond to less than half of the buffer,
we flush only the pages up to the \lastsortedzone{} and attempt to 
bulk load, if possible. That way, we avoid the sorting cost 
\emph{before} flushing. At the same
time, the rest of the pages are sorted and moved towards the left 
to make space for new inserts.

%

\subsection{Optimizing Read Queries}
\label{sec:w-r-opt-buf-tree}

While flushing the \pinBufName{} helps to harness the sortedness by
increasing the fraction of inserts that are bulk loaded, it has an adverse
impact on read performance. Specifically, every query goes through the following
steps: (i) search the buffer that may contain a sorted part
and an unsorted part, and (ii) perform a tree search. In the worst-case,
a read query will need a full scan of the buffer. We now discuss how
to reduce the cost of a lookup aiming to make it as close
as possible to that of the underlying tree.

\begin{figure}[t]
	\centering
    \includegraphics[scale=0.43]{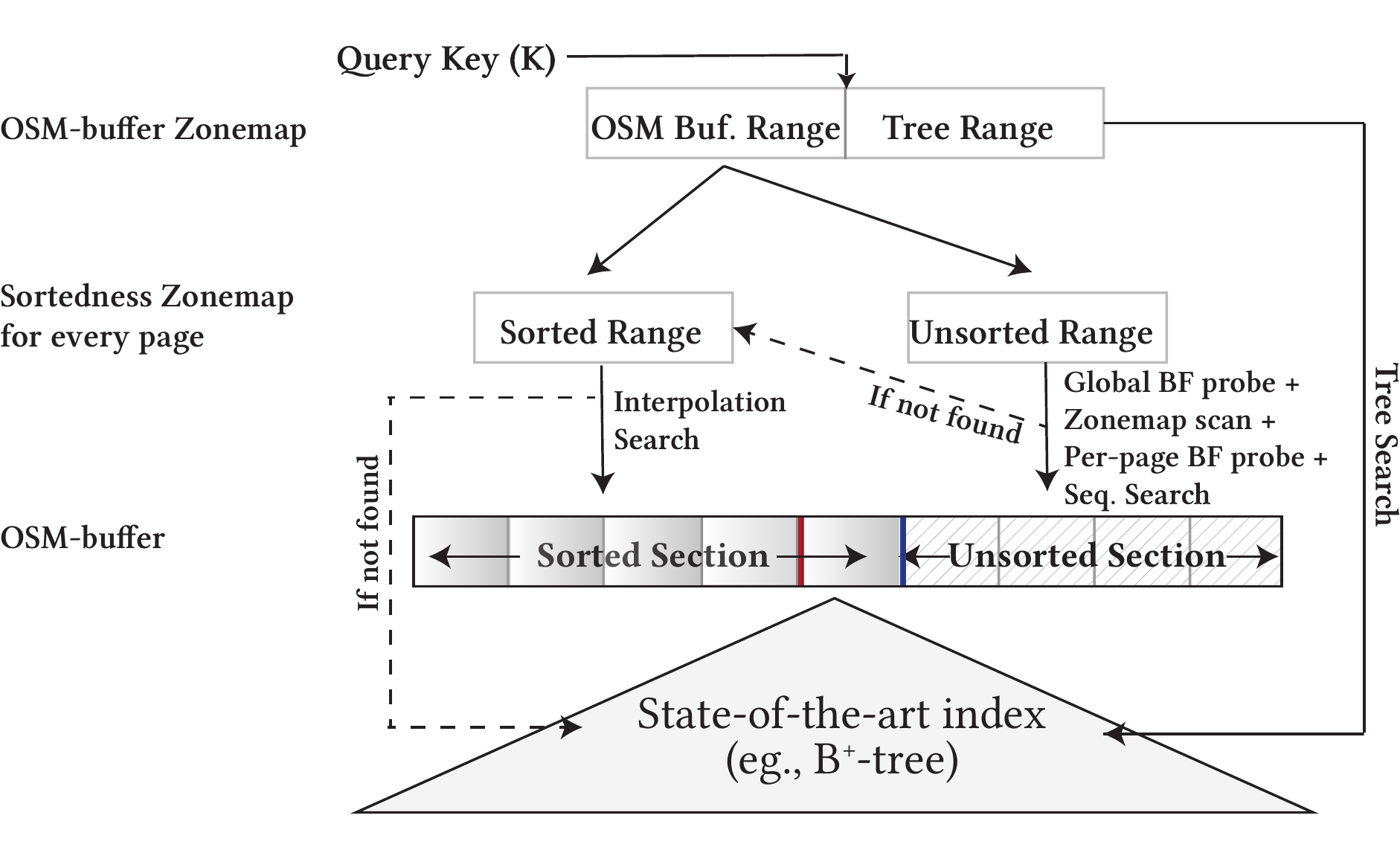}
\vspace{-0.15in}
	\caption{The lifecycle of a point query.}
\label{fig:osmBufPointQuery}
\vspace{-0.2in}
\end{figure}

\Paragraph{Scanning the Unsorted Section First}
In steady-state, the \pinBufName{} can be in one of two states:
(i) fully sorted or (ii) it can have a sorted portion and an unsorted 
portion. Note that even if the \lastsortedzone{} is moved 
to the beginning of the buffer, we still have half of the buffer sorted.
So, for any search query, we only need
to scan the unsorted portion of the buffer that contains the most 
recent data. If the lookup key is not found in this part of the buffer,
we continue to efficiently search the sorted section of the buffer, and if
the lookup has still not terminated, we search the tree. Note that if
the key is found in the buffer, we can terminate and
avoid searching the tree, as this
will be the most recent version of the key. The lifecycle
of a query is shown in Figure \ref{fig:osmBufPointQuery}.

\Paragraphit{A BF to Skip the Unsorted Section}
The unsorted section is up to half of the \pinBufName{} and, thus, 
holds a small fraction of the overall data (residing in the buffer
and the tree). As a result, most queries will not find the desired key 
in it. Hence, to avoid the cost of unnecessary scanning the unsorted section, we employ a BF that is continuously updated as new entries 
are inserted. This drastically reduces the cost of queries that do 
not terminate in the unsorted section. 

\Paragraphit{Using Zonemaps to Skip Pages in the Unsorted Section}
When the BF returns a positive result, all pages of
the unsorted section are marked for scanning. However, we can skip many unnecessary page accesses using the Zonemaps that are already part of the \pinBufName{} (used to 
identify the \lastsortedzone{}). 

\begin{figure}[t]
	\centering
    \includegraphics[scale=0.45]{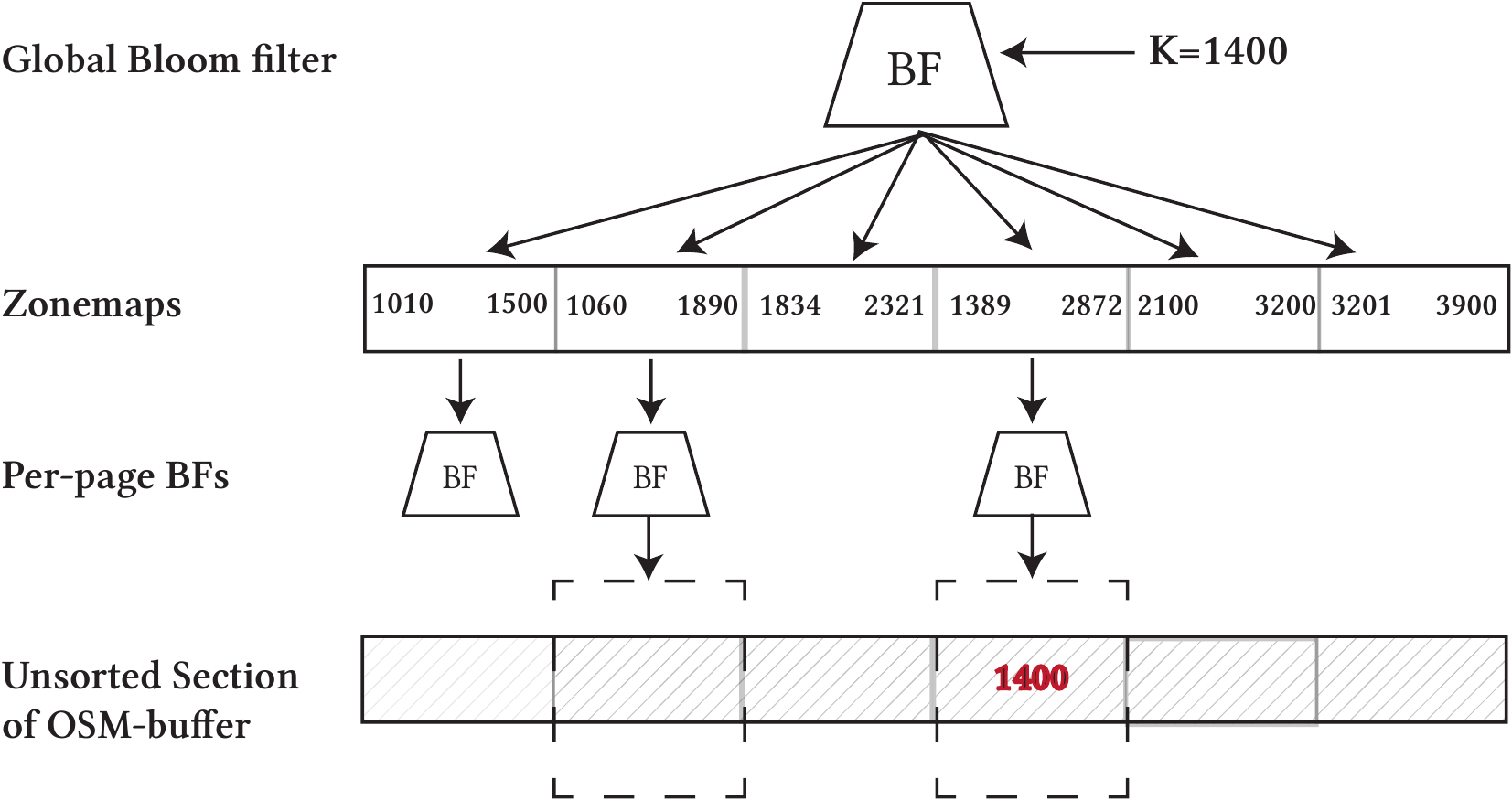}
	\caption{Focusing on the unsorted part of the buffer, a query will first visit the global BF. If the BF query is positive, it will ask all Zonemaps, and then the qualified per-page BFs before it scans any page.}
\label{fig:osmBufPointQueryUnsorted}
\vspace{-0.15in}
\end{figure}

\Paragraphit{Using Per-Page BF} While the BF mentioned above and the 
Zonemaps help avoid many unnecessary accesses, they are not
enough if the data is scrambled. Hence, we also maintain a BF per page, which
is updated as data is appended to the buffer. 
Overall, a query starts searching in the unsorted section by first visiting the 
\emph{global} BF (with respect to the unsorted section). As shown in 
Figure~\ref{fig:osmBufPointQueryUnsorted} for a search query on key 1400,
if the BF returns a positive result, we access all Zonemaps to find which 
pages contain the key in question. Subsequently, for the Zonemaps that contain
the key, we probe the per-page BF, and we visit only the qualifying pages. 

%

\Paragraph{Interpolation Search to Search Sorted Section}
After searching the unsorted part of the buffer is complete, if 
the query has not yet terminated, it will search the sorted section. Note
that after every flush the retained data is sorted. Irrespectively
to whether the newly inserted data overlap with the sorted section (and,
thus, move the \lastsortedzone{}), the data retained after the previous 
flush remain sorted, and we maintain the position in the buffer until
which the data is in sorted order, as \previousboundary{}. While 
the \lastsortedzone{} may move to the left as new entries are inserted 
into the buffer, the \previousboundary{} may only move rightwards
as long as entries are inserted in fully sorted order, and until the 
first out of order entry is inserted. Since the sorted section of
the buffer is a contiguous sorted array, we employ interpolation 
search~\cite{Perl1978,VanSandt2019} which finishes in
$O(log(log(N)))$ steps, a notable upgrade from the binary search and is efficient 
unless there is very high data skew, in which case simply binary search or 
a variation of 
exponential sorting~\cite{Bentley1976} can also be employed.

\Paragraph{An Optimized Read Query} Putting it all together, we have
now an optimized read query path that avoids the vast majority of 
unnecessary data accesses. As Figure \ref{fig:osmBufPointQuery} shows,
we maintain two more Zonemaps: one for the \pinBufName{} and one for the
tree. Hence, if the desired key is not in the range of the buffer, we can
skip the buffer entirely. On the other hand, in the worst case, we have
to access the unsorted section of the buffer. However, due to the per-page
BFs, even if the data is completely scrambled, we will only access a very
small number of pages from the unsorted section.

\subsection{Fine-tuning OSM-tree}
\label{sec:osm}

In the previous two subsections, we put together a new meta-design that
absorbs inserts faster if they come as near-sorted. Our design further keeps the
read query cost close to the cost of the underlying tree. This new
meta-design, however, introduces new components and a few tuning
knobs that can be further tuned. We begin by the fill and split factors
and the choice of sorting algorithms that affect mostly the insertion
latency, and we then discuss query-driven sorting to adaptively
accelerate read queries.

\Paragraph{Adjusting Fill \& Split Factor}
The textbook bulk loading algorithms used in  
Section \ref{sec:w-opt-buff-tree} fills every node with the bulk 
loaded data to maximize node utilization (and, thus, minimize space
amplification). Since we anticipate several of top-inserts (the fraction
of which depends on the sortedness of the workload), we also leave in
every bulk loaded node several empty slots to facilitate
top-inserts without expensive cascading splits. Hence, we adjust the
fill factor of every bulk loaded leaf to be 95\%. 

Similarly, in the textbook insertion and bulk loading, when an internal
node is full it splits right in the middle to generate two half-full
internal nodes. Since during bulk loading, we anticipate that most of
the future inserts will be of larger values, we also adjust the split 
factor to 80\%, as shown earlier in Figure \ref{fig:design-elements}(c). 
This allows us to maintain most of the internal nodes of the underlying
tree nearly full, even when the data is coming fully sorted, and 
essentially avoid the worst-case space amplification of \bplustree{s}
for sorted data. In addition to reducing space amplification, we also
reduce the total number of node splits, leading to lower overall insertion
cost. 


\Paragraph{Choice of Sorting Algorithm}
To reduce the cost of read queries, we sort the buffer
after every flush making sorting small data collections a very frequent 
operation. We consider three algorithms:
(i) \emph{quicksort}, because it is the most common algorithm and has minimal space requirements, 
(ii) \emph{$(K,L)$-adaptive sorting} \cite{Ben-Moshe2011}, because it aggressively takes into account pre-existing data sortedness (at the expense of $O(K+L)$ space usage), and 
(iii) \emph{mergesort}, because it maintains relative order of duplicate values (a key property to know which overlapping entry is the latest one) at the expense of $O(n)$ space usage. 
Because we need to maintain relative order of duplicates we are constrained between mergesort and $(K,L)$-adaptive sorting. Our analysis
shows that for low data-sortedness mergesort outperforms $(K,L)$-adaptive sorting (in fact, $(K,L)$-adaptive sorting fails for very high values of $K$ or $L$). However, for $K<10\%$ or $L<5\%$, their performance is similar and we opt
for $(K,L)$-adaptive sorting because it has smaller space requirements $(K+L<n)$. In summary,
when the estimated values of $K$ and $L$ are
$K<10\%$ or $L<5\%$ of the buffer size we employ
$(K,L)$-adaptive sorting, and otherwise mergesort,
and specifically the C++ standard library 
implementation of \texttt{std::stable\_sort}.

\begin{figure}[t]
	\centering
    \includegraphics[width=0.65\columnwidth]{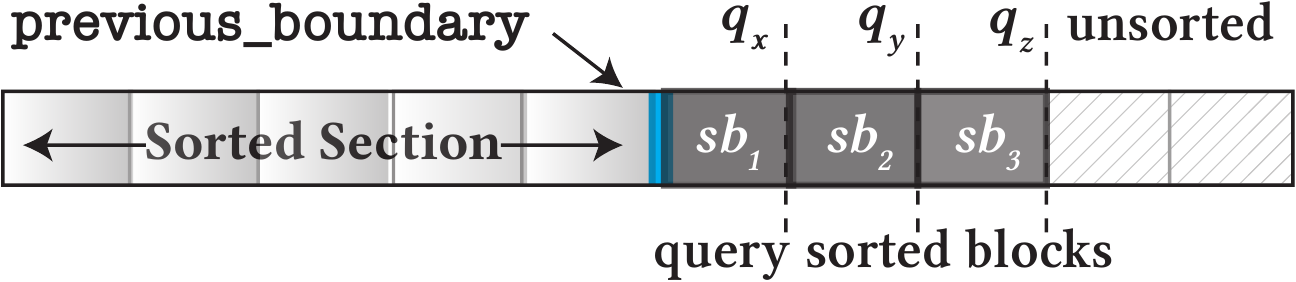}
	\vspace{-0.12in}
    \caption{As inserted data accumulate in the unsorted buffer above a threshold, an incoming query sorts this part as a standalone sorted component, to benefit future queries.}
    \label{fig:subsorted}
	\vspace{-0.2in}
\end{figure}

\Paragraph{Query-Driven Sorted Components}
In our current design, only the first part of the buffer that is sorted
can benefit from interpolation search. An additional read optimization
is to adaptively add structure to the unsorted part of the buffer
with incoming queries. 
We set a threshold of how large we allow the unsorted portion of the buffer to
be (\texttt{unsorted\_threshold}). When the threshold is exceeded, the next read query 
will 
sort this portion and create a new sorted component. Similar to progressive 
indexing~\cite{Holanda2019} that allocates a small indexing budget for 
every query, we allocate a small sorting budget for every query as long as 
we have enough entries in the unsorted component.
In the general case, the
\pinBufName{} may contain the main sorted section, multiple sorted 
components of size equal to \texttt{unsorted\_threshold}, and a small
unsorted section. For example, if the \texttt{unsorted\_threshold} is 10\%
of the buffer size, the buffer will contain five sorted runs (the sorted
section and four sorted components) and one unsorted section. 
The unsorted section still uses all the metadata 
discussed in \S\ref{sec:w-r-opt-buf-tree} and each sorted
component employs interpolation search to accelerate queries.


\Paragraph{\pinBufName{} Size}
The final tuning knob is the size of the \pinBufName{}. The goal is to
have a large enough buffer that can capture sortedness, focusing on
$L$, i.e., the maximum displacement from the expected position. 
On the other hand, a large buffer would negate any optimization for read
queries since the cost of scanning the unsorted part will dominate. 
In Section~\ref{sec:evaluation}, we vary the buffer
size and the relative values of $K$ and $L$, 
and we show that even with a buffer significantly smaller than $L$ we
can absorb sortedness to a large degree, without hurting read queries.

\vspace{-0.05in}
\subsection{Discussion}
\vspace{-0.02in}

\Paragraph{Handling Strings} 
Near-sortedness manifests typically through integers than any other data type, thus, our work focuses on integer representations of keys. However, the OSM-design paradigm and \sysName{} can be extended to other data formats such as strings.
Specifically, to address storage constraints, strings with variable sizes can be mapped to fixed-length representations using a dictionary. This accommodates all strings regardless of length. Subsequently, the fixed-length representations can be binary encoded before insertion into the index. We can ensure that the order is preserved through the mapping and the encoding steps. 
We aim to extend this work and further include a study over \emph{near-sorted strings} including designing a synthetic workload generator for near-sorted strings, which to our knowledge has been unexplored.

\Paragraph{Learned Indexes} 
Contrary to traditional index data structures, learned indexes, use machine learning to learn a model reflecting patterns in the data and enable automatic synthesis of indexes at a low engineering cost \cite{Kraska2018}. 
Lookups in learned indexes avoid expensive tree traversals, aiming to offer lower access latency. 
However, existing learned index proposals require trading workload generality for accuracy. Specifically, to achieve sufficient lookup accuracy, learned indexes make the assumption of completely sorted keys, thus, depend on prior knowledge of the data.
\sysName{} addresses the cost of tree traversal by adapting operations to the data. By taking advantage of inherent sortedness, \sysName{} significantly reduces the ingestion cost while offering low latency and full accuracy.
As future work, we further aim to extend existing learned indexes to make them sortedness-aware and relax the assumptions regarding input data.

\section{Data Sortedness Benchmark}
\label{sec:sortednessbench}
We present the \emph{data sortedness benchmark} for testing indexes against varying sortedness. The benchmark uses the $(K,L)$-near sorted 
metric (discussed in \S\ref{sec:background}) and is used in our evaluation
(\S\ref{sec:evaluation}). 


\begin{figure}[t]
    \centering
    \begin{subfigure}[b]{0.15\textwidth}
        \centering
        \includegraphics[scale=0.24]{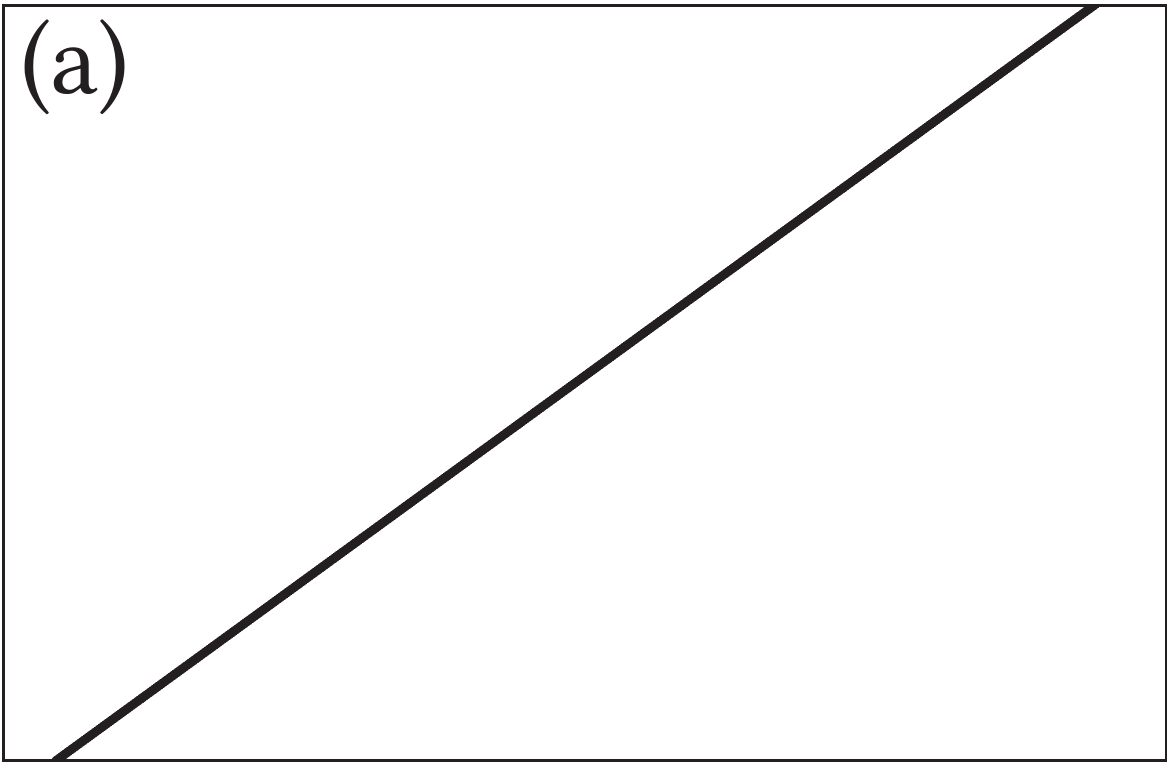}
    \end{subfigure}
    \hfill
    \begin{subfigure}[b]{0.15\textwidth}
        \centering
        \includegraphics[scale=0.24]{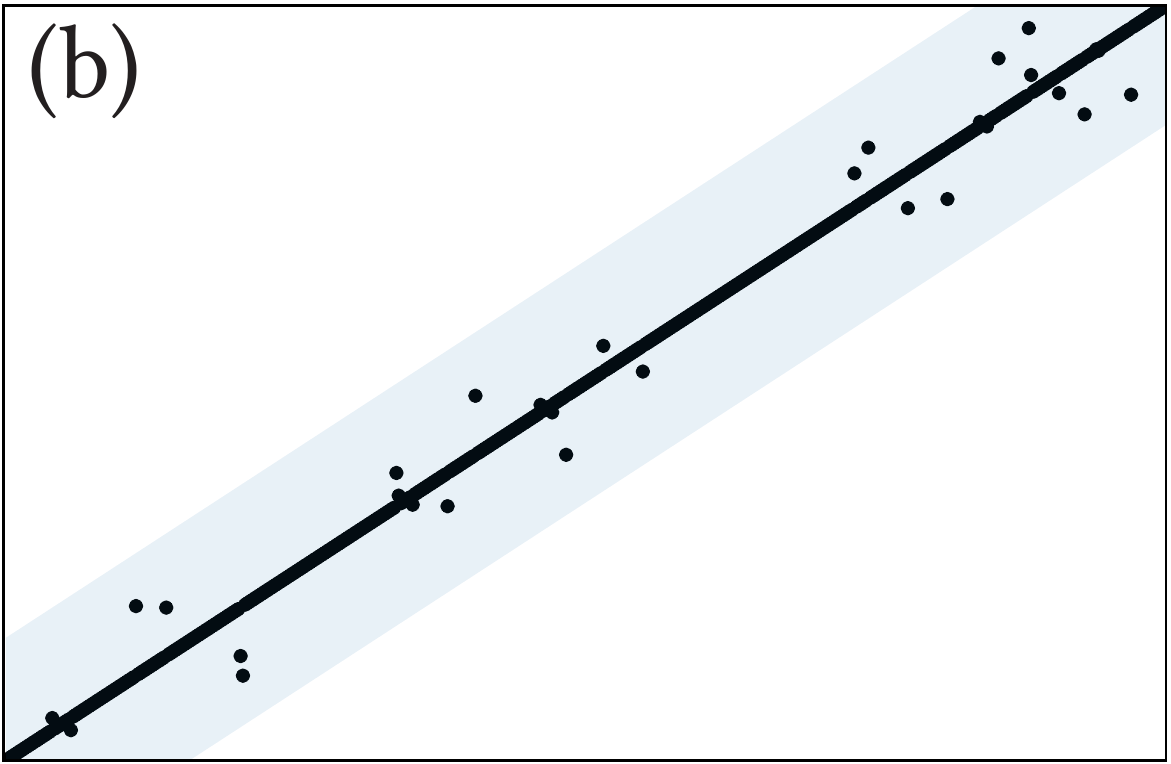}
    \end{subfigure}
    \hfill
    \begin{subfigure}[b]{0.15\textwidth}
        \centering
        \includegraphics[scale=0.24]{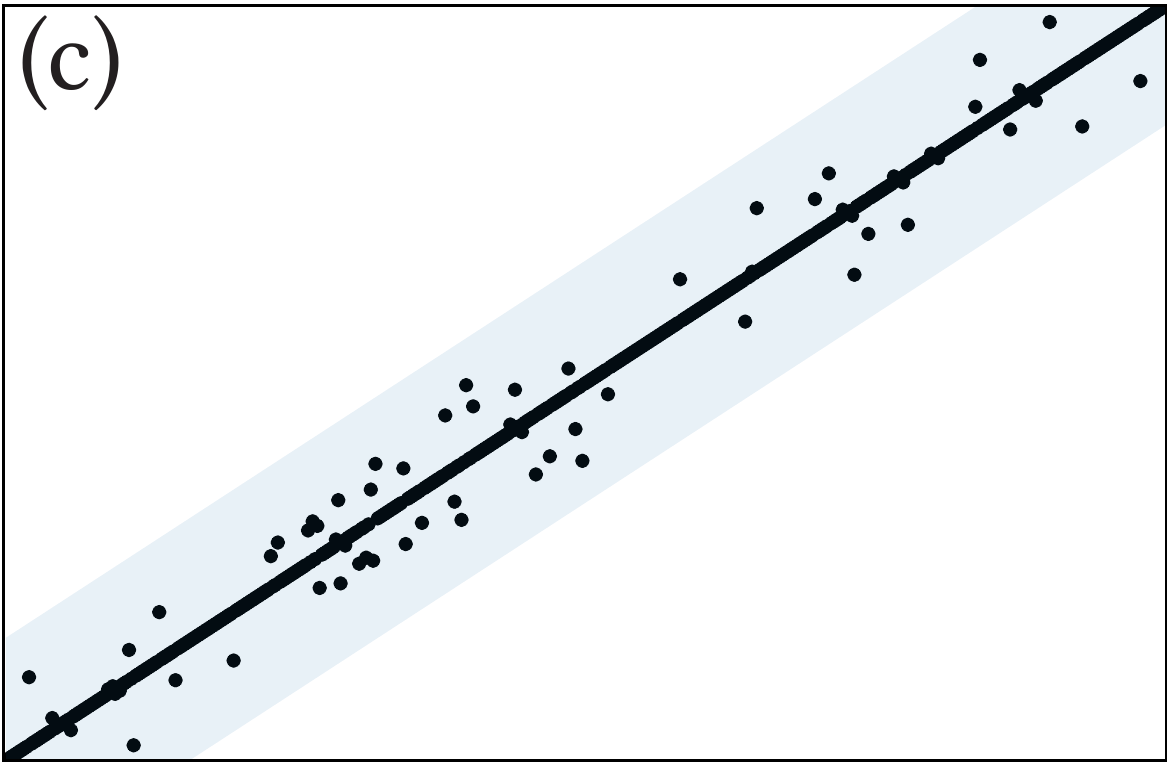}
    \end{subfigure}
    \begin{subfigure}[b]{0.15\textwidth}
        \centering
        \includegraphics[scale=0.24]{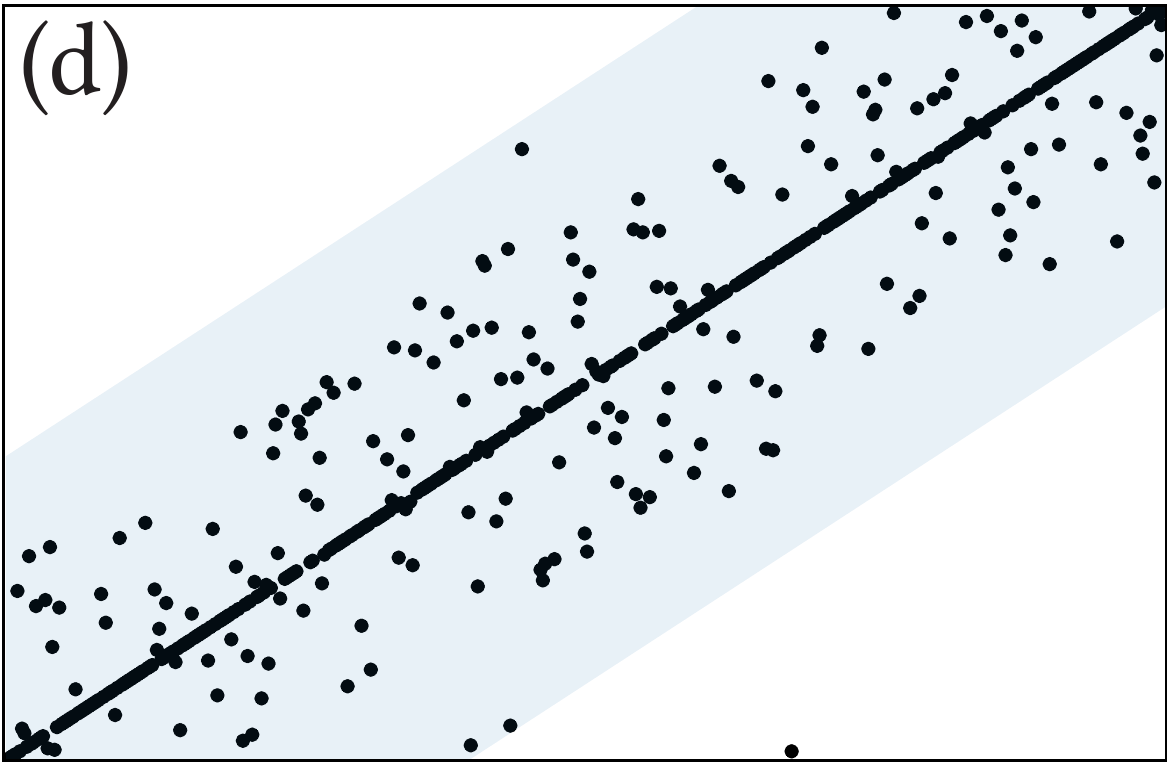}
    \end{subfigure}
    \hfill
    \begin{subfigure}[b]{0.15\textwidth}
        \centering
        \includegraphics[scale=0.24]{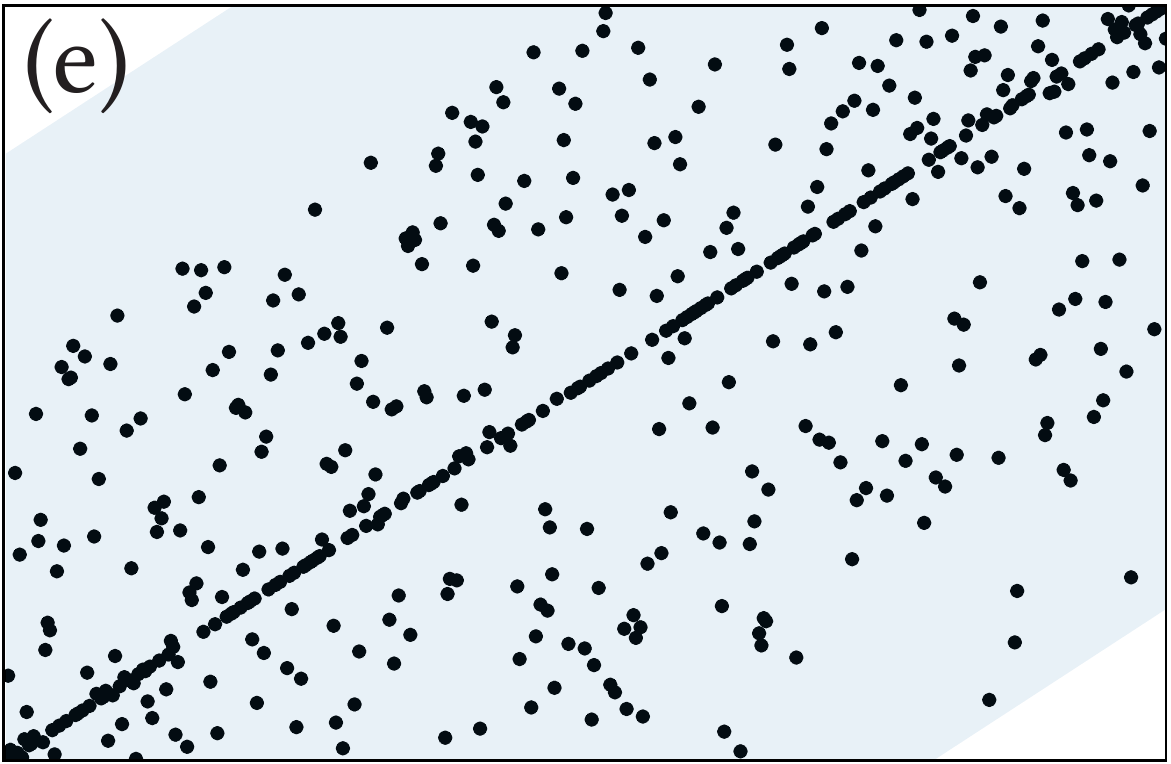}
    \end{subfigure}
    \hfill
    \begin{subfigure}[b]{0.15\textwidth}
        \centering
        \includegraphics[scale=0.24]{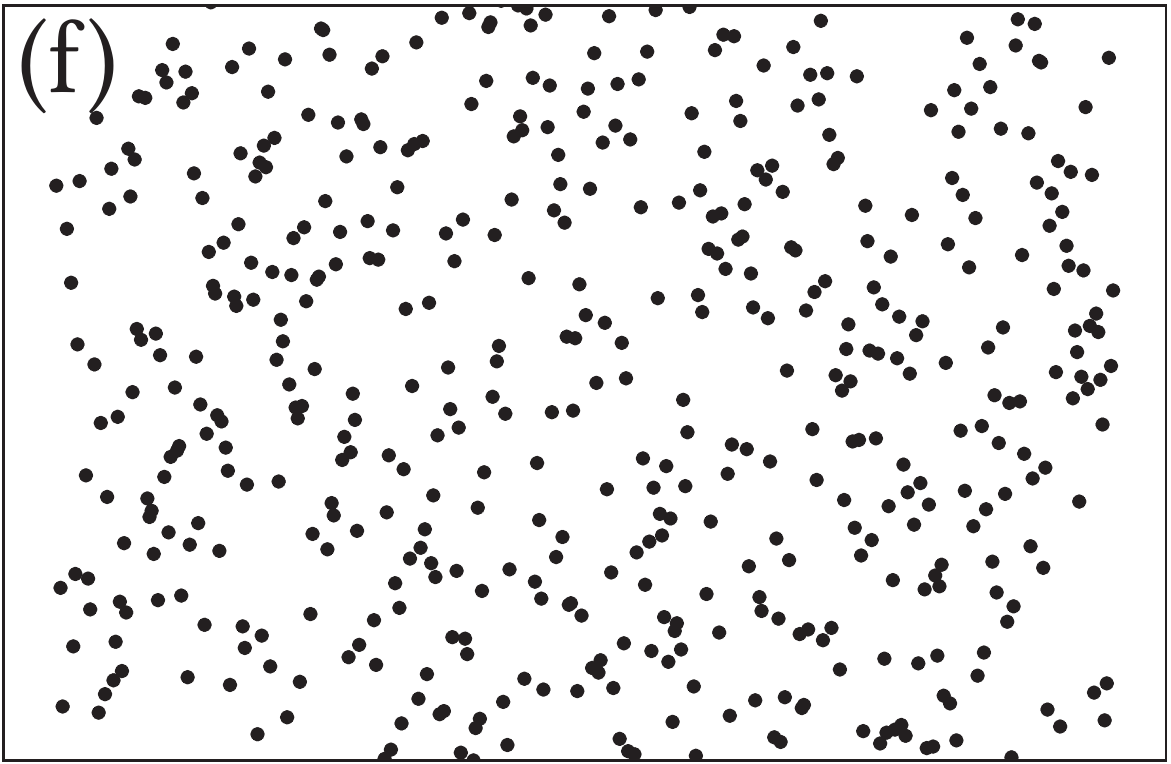}
    \end{subfigure}    
\caption{Workloads with varying degrees of \emph{sortedness}: 
	(a) sorted, (b) K=5\%, L=10\%, (c) K=10\%, L=10\%, 
	(d) K=25\%, L=25\%, (e) K=50\%, L=50\%, and (f)~scrambled (uni-rand). 
	x-axis: position of entry in data, y-axis: value of entry.}
\label{fig:workloads}
\vspace{-0.15in}
\end{figure}

\Paragraph{Benchmark Data}
The benchmark creates a family of differently sorted collections 
that vary in both $K$ and $L$ (as a fraction of the total data size). 
These metrics effectively capture sortedness by representing how many entries 
are out-of-order and how far apart are the entries from their actual position, underlining the 
effort it would take to establish order in the data collection. 
When $K=0\%$ or $L=0\%$ the dataset is fully sorted.
A dataset with $K=10\%$ and $L=2\%$ will have $10\%$ of the total 
entries out of order, each placed within a distance of $2\%$ of the total entries from its 
in-order position. 
Figure ~\ref{fig:workloads} shows a sample set of 
differently sorted collections. The x-axis denotes the position of the entry in a data collection
and the 
y-axis represents the value of the entry. The band across every illustration highlights the 
$L$-window through which elements can be shuffled. As this window grows, the 
band increases in width, meaning elements may be further apart from their ideal positions.

\Paragraph{Evaluation Metrics}
The sortedness benchmark evaluates the performance of a data structure in the presence of
variable sortedness by measuring: (i) ingestion performance, (ii) overall performance of a mixed 
workload with variable read/write ratio. 
The default settings for the benchmark include comparison with multiple 
$L$-windows ($1\%$, $5\%$, $10\%$, $25\%$, $50\%$), and various $K$ 
values for each window ($0\%$, $1\%$, $5\%$, $10\%$, $15\%$, $25\%$, $50\%$).

\Paragraphit{Ingestion Speedup}
This metric quantifies the benefit in terms of ingestion latency by comparing the 
underlying index with its OSM counterpart. The benchmark reports both raw performance 
numbers and the speedup that quantifies the ingestion benefit. 

\Paragraphit{Overall Speedup}
This metric quantifies the benefit when running a mixed workload with a varying degree 
of insert-to-lookup ratio
between $10\%$$:$$90\%$ and $90\%$$:$$10\%$. The benchmark reports raw performance and 
the overall speedup when comparing the underlying index with its OSM counterpart.

\Paragraphit{Microbenchmark Varying $K$ and $L$}
The above measurements are taken for fixed value of $K$ and $L$. The last set of measurements
of the sortedness benchmark vary both $K$ and $L$ within a window in order to capture
their impact. The reported values are raw performance and
speedup as well. 

%


\section{Experimental Evaluation}
\label{sec:evaluation}
We now present the experimental evaluation of \sysName{}.


\Paragraph{Experimental Setup}
We run the experiments in our in-house server equipped with 
two sockets each with an Intel Xeon Gold 5230 2.1GHz processor with 20 cores and virtualization 
enabled. The server has 384GB of RDIMM main memory at 2933 MHz with 27.5MB L3 cache and a 240GB SSD. The machine runs on CentOS 8. 

\Paragraph{Index Design}
We use a state-of-the-art \bplustree{} implementation \cite{Bingmann2007} and build on top of it 
to add support for in-order bulk insertion. Our \bplustree{}  implementation is equipped with
a bufferpool of 300GB, so all our experiments are purely 
in-memory.
Note that the bufferpool is 
orthogonal to the \pinBufName{}, which does not have a disk-resident counterpart.
The indexed key and payload,
unless otherwise noted, is 8B in total, with keys and values of 4B each, and by default,
we use 4KB index pages.


\Paragraph{Default Setup}
The default size of the \pinBufName{} is 40MB which can hold up to 5M entires. The \pinBufName{}
is essentially a dense array accompanied by the Zonemaps and Bloom filters as discussed in
Section \ref{sec:design}. For the BFs, we use $10$ bits-per-entry and maintain the filters at two 
granularities: (i) one for the entire \pinBufName{} and (ii) one for each page in the \pinBufName{}. 
The BFs use \emph{MurmurHash}~\cite{Appleby2011} for hashing, following 
the state-of-the-art \cite{Zhu2021}.


\Paragraph{Workload}
Given the lack of sortedness benchmarks, we create our custom 
workload generator
based on 
the $(K,L)$-nearly sorted metric~\cite{Ben-Moshe2011} to generate workloads with varying degrees of 
sortedness as discussed in Section \ref{sec:sortednessbench}. 
Unless otherwise mentioned, the ingestion workload consists of $500$M key-value entries with a total 
size of $4$GB, and the query workload has a variable number of uniform random non-empty point lookups, interleaved with inserts after $80\%$ of the ingestion is complete. 

\Paragraph{OSM Tuning}
Unless otherwise mentioned, we tune our \sysName{} as
follows. The \pinBufName{} flushes $50\%$ of the entries when saturated. The 
nodes split as $80$$:$$20$ between the left and right one, and the in-order bulk insertion fills 
every leaf up to $95\%$. For workloads with queries interleaved with inserts, a sorted block is 
created (triggered by a query) after at least $500$K new entries ($10\%$ of the buffer capacity) are 
accumulated.

\subsection{Mixed Workload}
\label{subsec:mixed_workload}
We first compare the performance of \sysName{} with \bplustree{} by executing a set of mixed 
workloads with interleaved inserts and queries. We vary the read-write ratio, constructing a 
continuum between a write-heavy and a read-heavy workload. For each workload, we also 
vary the sortedness for the ingested data as: (i) \emph{fully sorted}, 
(ii) \emph{near-sorted} (K=5\%, L=5\%), (iii) \emph{less sorted} (K=50\%, L=50\%), and (iv) \emph{scrambled} (uniformly 
random).
For each experiment, we measure the speedup offered by \sysName{} as 
$speedup = \frac{latency(\text{\textit{\bplustree{}}})}{latency(\text{\textit{\sysName{}}})}$, where 
$latency(\text{\textit{\bplustree{}})}$ and $latency(\text{\textit{\sysName{}}})$ refer to the total workload execution latency of the \sysName{} and \bplustree{}, respectively.

\begin{figure}[!tb]
    \centering
    \includegraphics[width=\columnwidth]{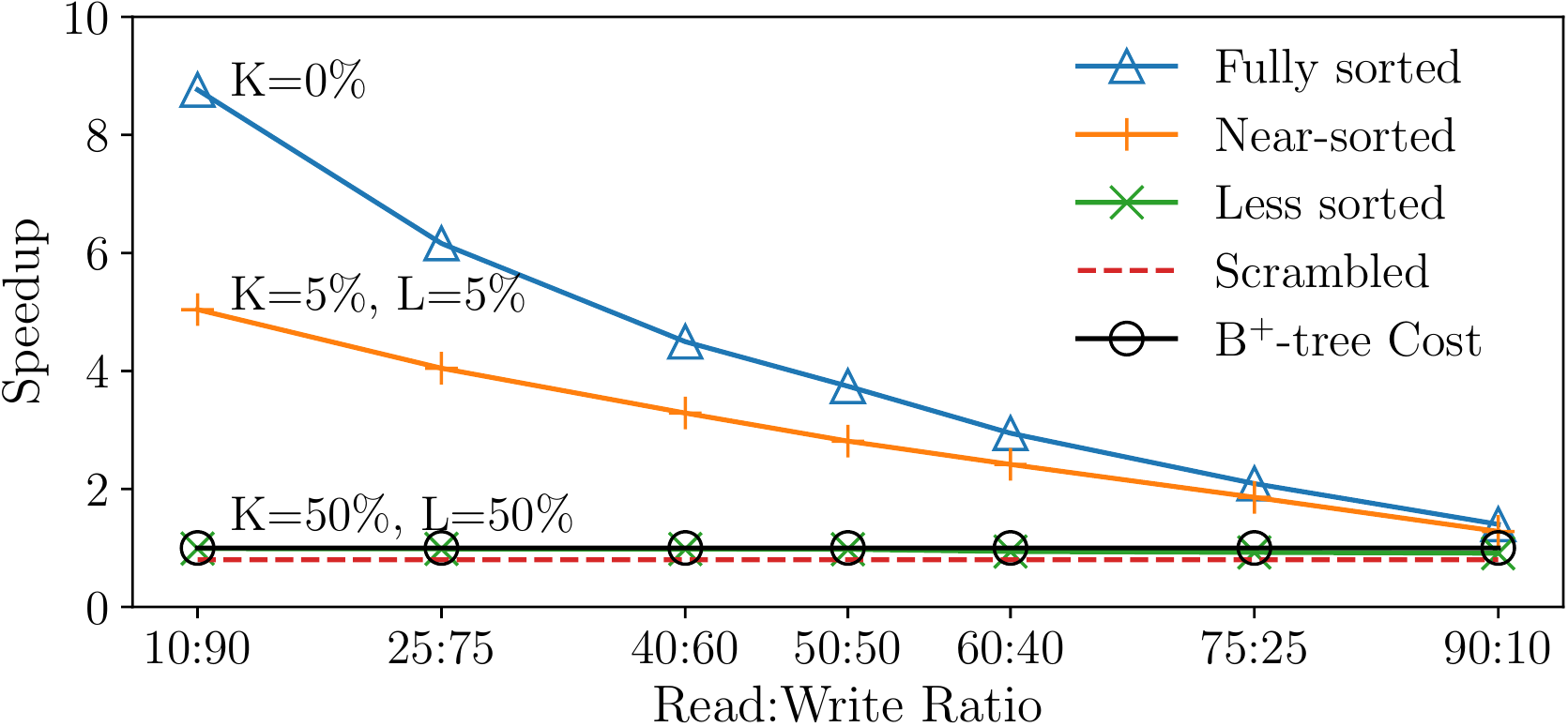}
    \caption{\sysName{} (Buffer size$=$$5$M) is efficient with reasonable data sortedness for any 
    read-write ratio.}
    \label{fig:mixed_perf}
    \vspace{-0.2in}
\end{figure}

\Paragraph{\textbf{\sysName{} Outperforms B$^+$-Tree}}
Figure ~\ref{fig:mixed_perf} shows that \sysName{} significantly outperforms 
\bplustree{} if the data is fully sorted or near-sorted. 
For an ingestion-heavy workload, \sysName{} leads to $8.8\times$ speedup for fully sorted data and $5\times$ better for near-sorted data in ingestion-heavy workloads. 
\setlength{\columnsep}{13pt}
\setlength{\intextsep}{10pt}
\begin{wrapfigure}{r}[-0.1in]{0.21\textwidth}
	\vspace{-0.1in}
	\centering
	\hspace{-0.5in}
	\includegraphics[scale=0.3]{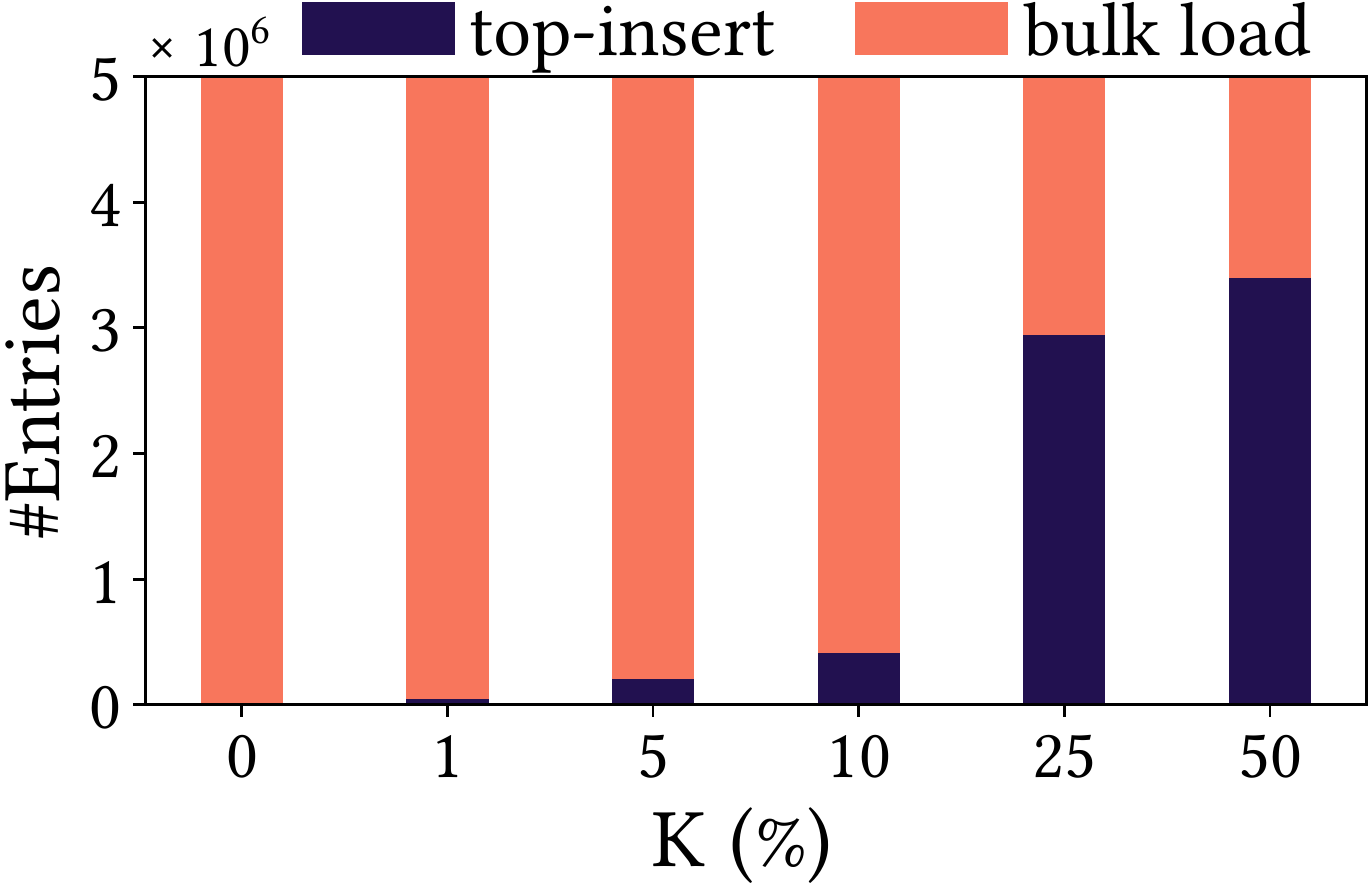}
	\vspace{-0.12in}
	\hspace{-0.5in}
	\caption{For higher $K\%$, \sysName{} performs more top-inserts and bulk loads fewer entries.}
	\label{fig:breakdown}
	\vspace{-0.15in}
\end{wrapfigure}
\sysName{} achieves this by buffering entries in-memory to add structure to the data and reduce the number of top-inserts. 
In the case of fully sorted data, all of the entries are ingested through the bulk insertion, while for the 
near-sorted data, only $\sim$$4\%$ of the entries are top-inserts and the remaining follow bulk insertion. 
Figure \ref{fig:breakdown} shows that \sysName{} performs significantly fewer top-inserts for workloads with a high degree of sortedness. 
For fully sorted or nearly sorted workloads (i.e., $K$ or $L \leq 10\%$), \sysName{} ingests $>$$90\%$ data through bulk loading, and thereby, reduces the overall cost for ingestion.
\begin{figure*}[t]
    \centering
    \begin{subfigure}[t]{0.25\textwidth}
        \centering
        \includegraphics[scale=0.315]{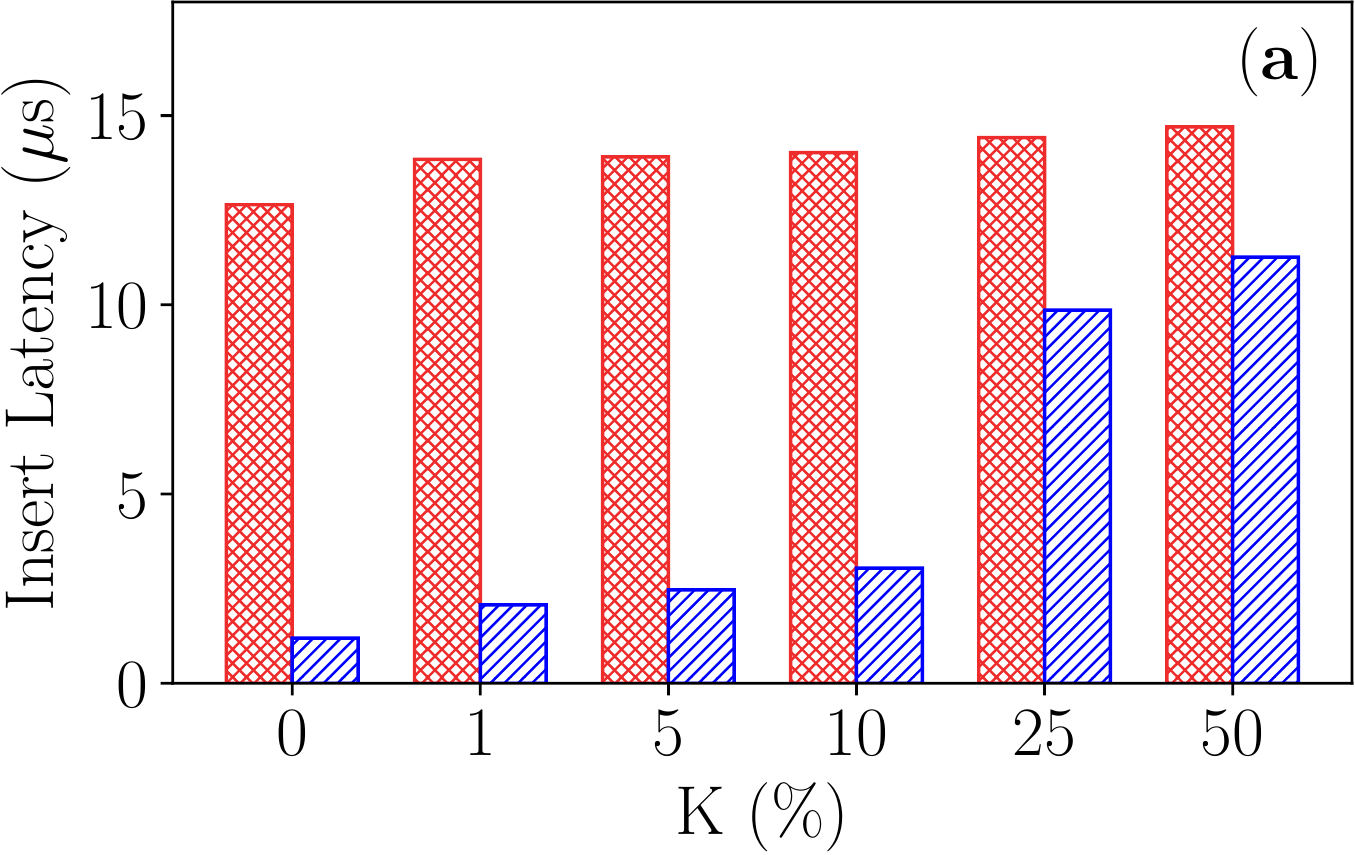}
    \end{subfigure}
	\hspace{-2mm}
    \begin{subfigure}[t]{0.25\textwidth}
        \centering
        \includegraphics[scale=0.315]{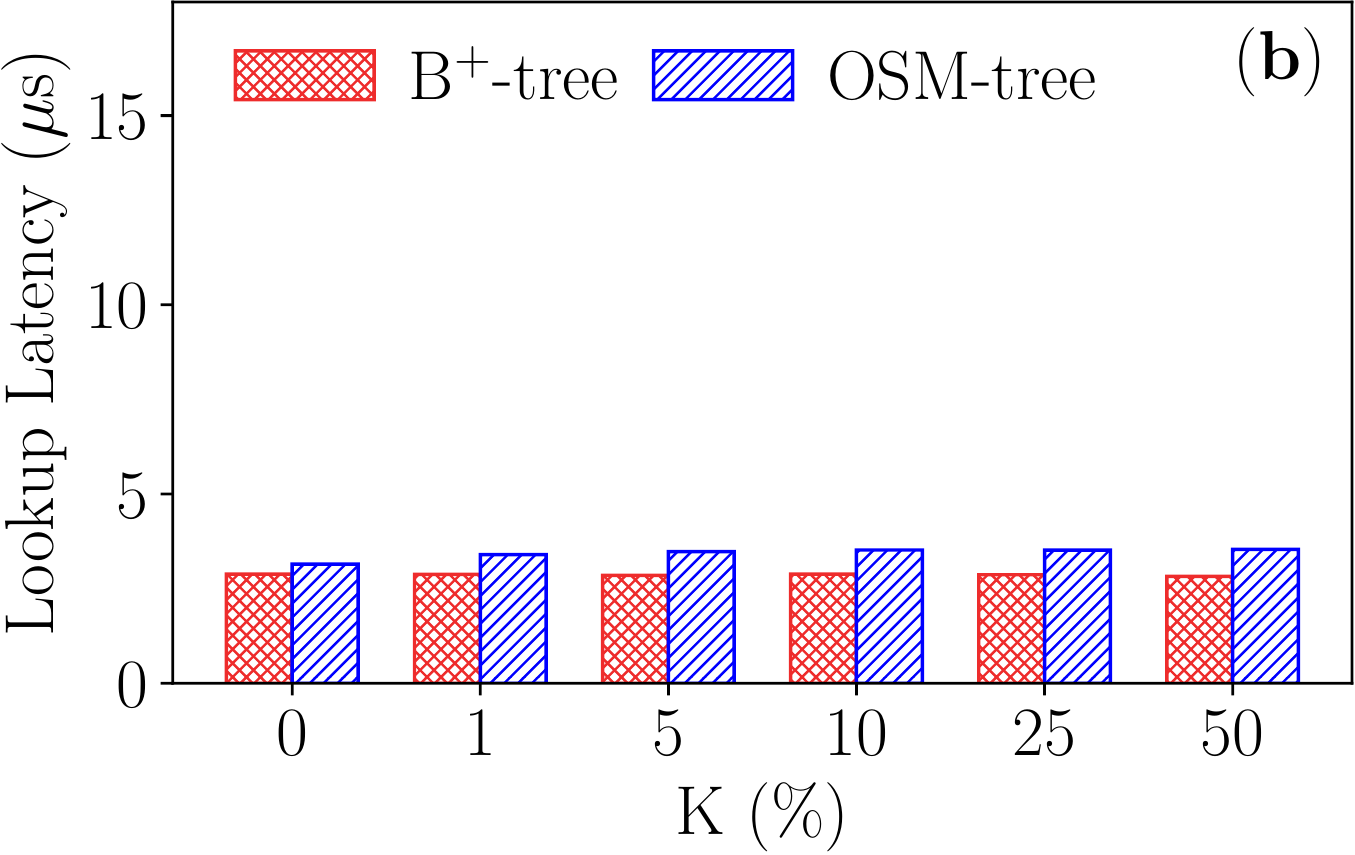}
    \end{subfigure}
	\hspace{-2mm}
    \begin{subfigure}[t]{0.25\textwidth}
        \centering
        \includegraphics[scale=0.315]{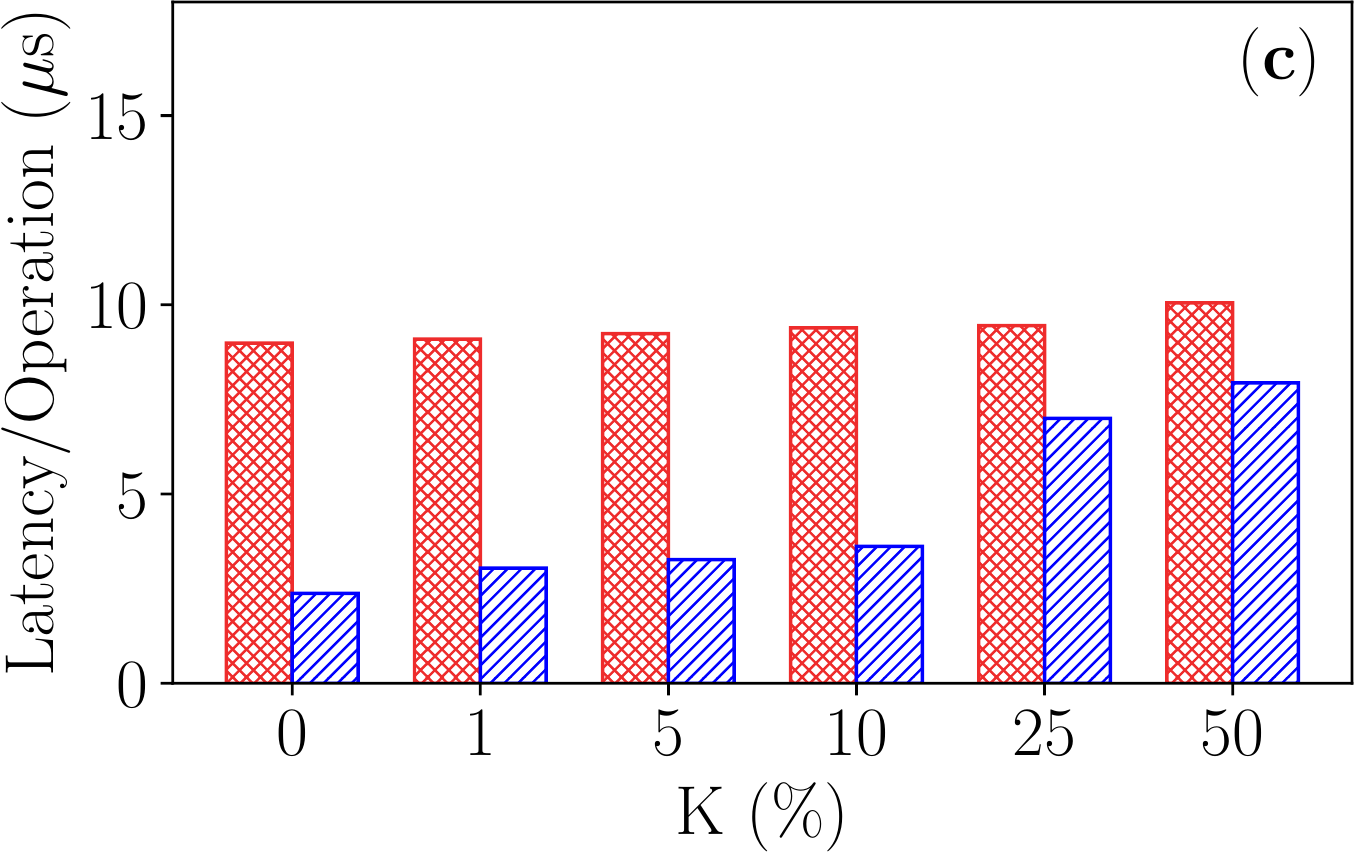}
    \end{subfigure}
	\hspace{-2mm}
    \begin{subfigure}[t]{0.25\textwidth}
        \centering
        \includegraphics[scale=0.315]{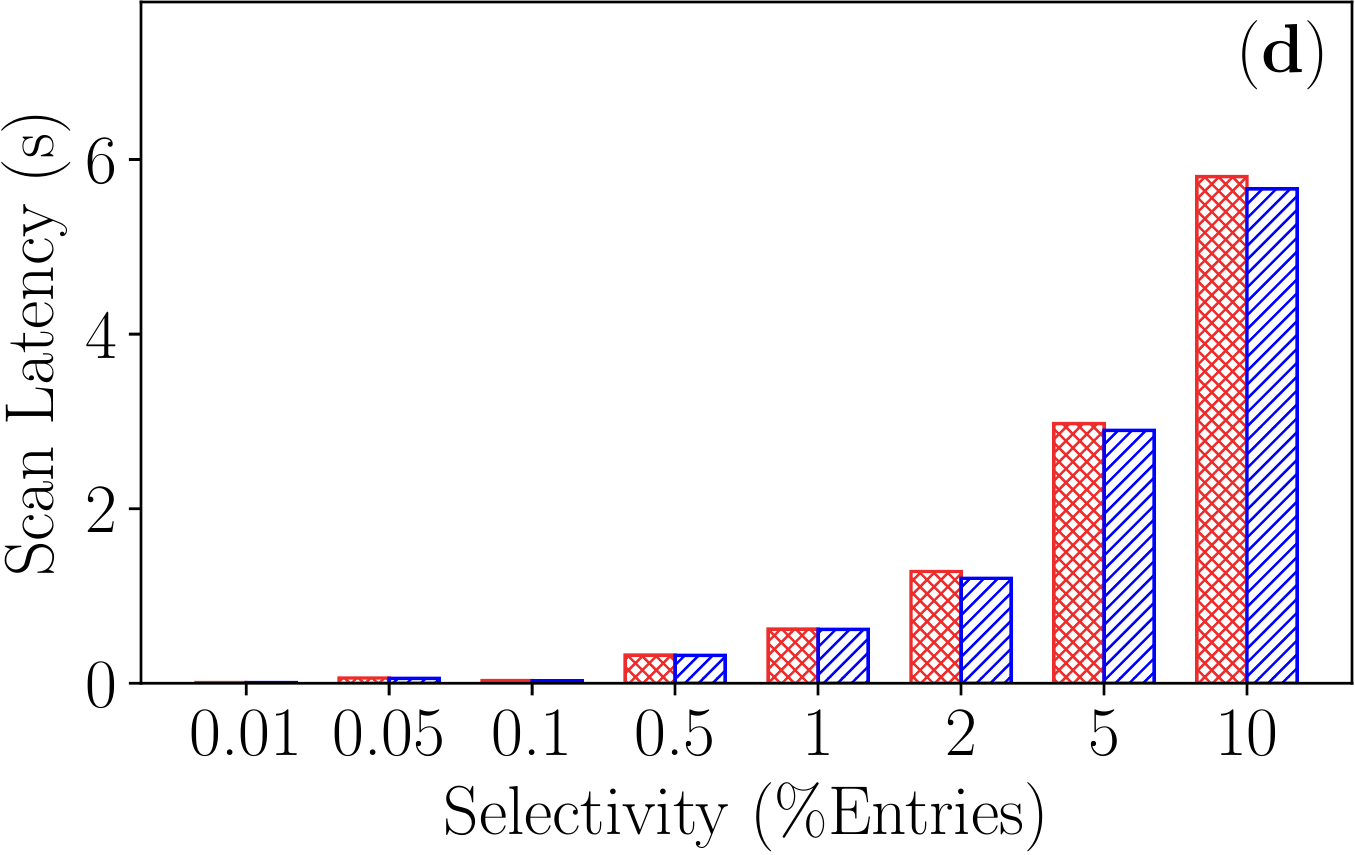}
    \end{subfigure}
	\vspace{-0.05in}
    \caption{Performance of \sysName{} (buffer size$=$$40$MB) for $L$$=$$5\%$. (a) \sysName{} offers better ingestion performance with workloads with some degree of data sortedness. (b) \sysName{} incurs a small overhead for point queries. (c) For a mixed workload with equal proportions of reads and writes, the ingestion-benefits outweigh the lookup-overhead and offers better overall performance. (d) \sysName{} offers competitive performance for both short and long range scans.}
    \label{fig:raw_perf}
	\vspace{-0.05in}
\end{figure*}
As data becomes less sorted, \sysName{} mimics the behavior of a \bplustree{}. This is because, as the degree of data sortedness decreases, \sysName{}'s ability to capture the out-of-order elements using a relatively small buffer ($1\%$ of data size) diminishes, and the number of top-inserts performed becomes comparable to that of a \bplustree{}.

%
Regardless of data sortedness, the benefit
of \sysName{} is more pronounced for
write-intensive workloads.
Conversely, in Figure ~\ref{fig:mixed_perf} we observe that for a lookup-heavy workload ($90\%$ lookups), 
\sysName{} offers a speedup of $1.4\times$ and $1.3\times$ for fully sorted and nearly sorted data, respectively. This is because the 
significant performance benefits of \sysName{} during ingestion 
are countered by the lookup overhead incurred. 

\Paragraph{Ingesting Scrambled Data Does Not Benefit from \sysName{}}
When the ingestion is scrambled, \sysName{} does not offer
performance benefits. Specifically, when the data is 
generated uniformly random, using a \bplustree{} is about 20\% 
faster than \sysName{}, regardless of the proportion of 
ingestion and lookups in a workload. This is attributed to
the fact that a finite buffer 
is unable to capture the (minimal) sortedness of the 
incoming data. This, 
in turn, forces \sysName{} to always perform top-inserts. 
For that reason, the OSM buffer management cost (sorting the 
buffer, managing metadata, and probing BFs during lookups) does 
not pay off, however, it keeps the penalty to a modest 20\%. 
This observation is inline with our goals and our
expectation. While \sysName{} is very useful for a varying
degree of sortedness, for fully scrambled data, the worst-case
guarantees of a classical \bplustree{} are enough.

\Paragraph{\sysName{} Reduces Memory Footprint for High Data Sortedness}
The memory footprint of \sysName{} is $0.52\times$ ($0.6\times$) of
the size of a \bplustree{} for fully sorted (near-sorted) data as shown in Table \ref{tab:memory_footprint}.
The gain in memory footprint is attributed to the 80:20 split ratio of \sysName{} nodes, which achieves higher fill factor for high data sortedness.
On the other hand, for less-sorted data the 80:20
split ratio does not offer any benefit.

\subsection{Raw Performance} \label{subsec:raw_perf}
Next, we compare the \sysName{} and \bplustree{} in terms of 
ingestion and query performance separately. We vary the number of out-of-order entries 
($K\%$) in the ingestion workload, while keeping the maximum 
displacement of an out-of-order entry ($L\%$) constant. We 
keep $L$ constant because capturing the degree of sortedness with 
different $L$ values requires changing the buffer size as a 
function of $L$. In Sections~\ref{subsec:workload} 
and~\ref{subsec:osm_tuning}, we discuss the 
implications on performance of varying both $K$, $L$ as well
as the buffer size.

\Paragraphit{Setup}
To measure the raw ingestion performance, we first ingest $500$M entries ($4$GB), and for the query performance, 
we perform $50$M ($10\%$) lookups on the inserted keys. 
We report the worst-case lookup performance, by making sure
that the buffer is full before executing any query. 
To analyze the range scan performance, we execute $100$ range scans generated randomly from the key-domain on a preloaded database for different scan-selectivities.

\begin{figure}[t]
    \centering
	\hspace{1mm}
    \begin{subfigure}[t]{0.2\textwidth}
        \centering
        \hspace{-0.4in}
		\includegraphics[scale=0.24]{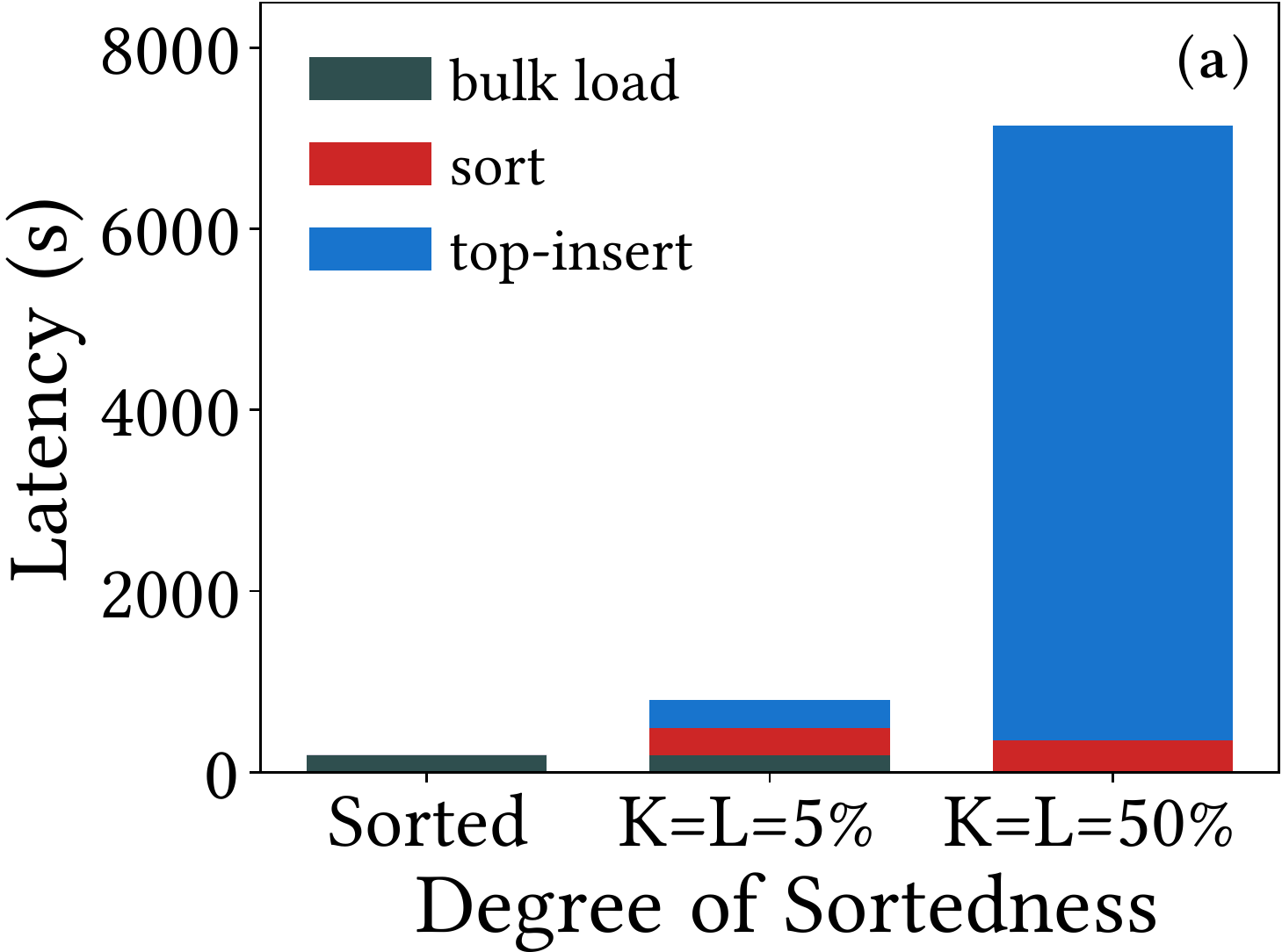}
		\label{}
    \end{subfigure}
	\hspace{-3mm}
    \begin{subfigure}[t]{0.2\textwidth}
      \centering
		\includegraphics[scale=0.24]{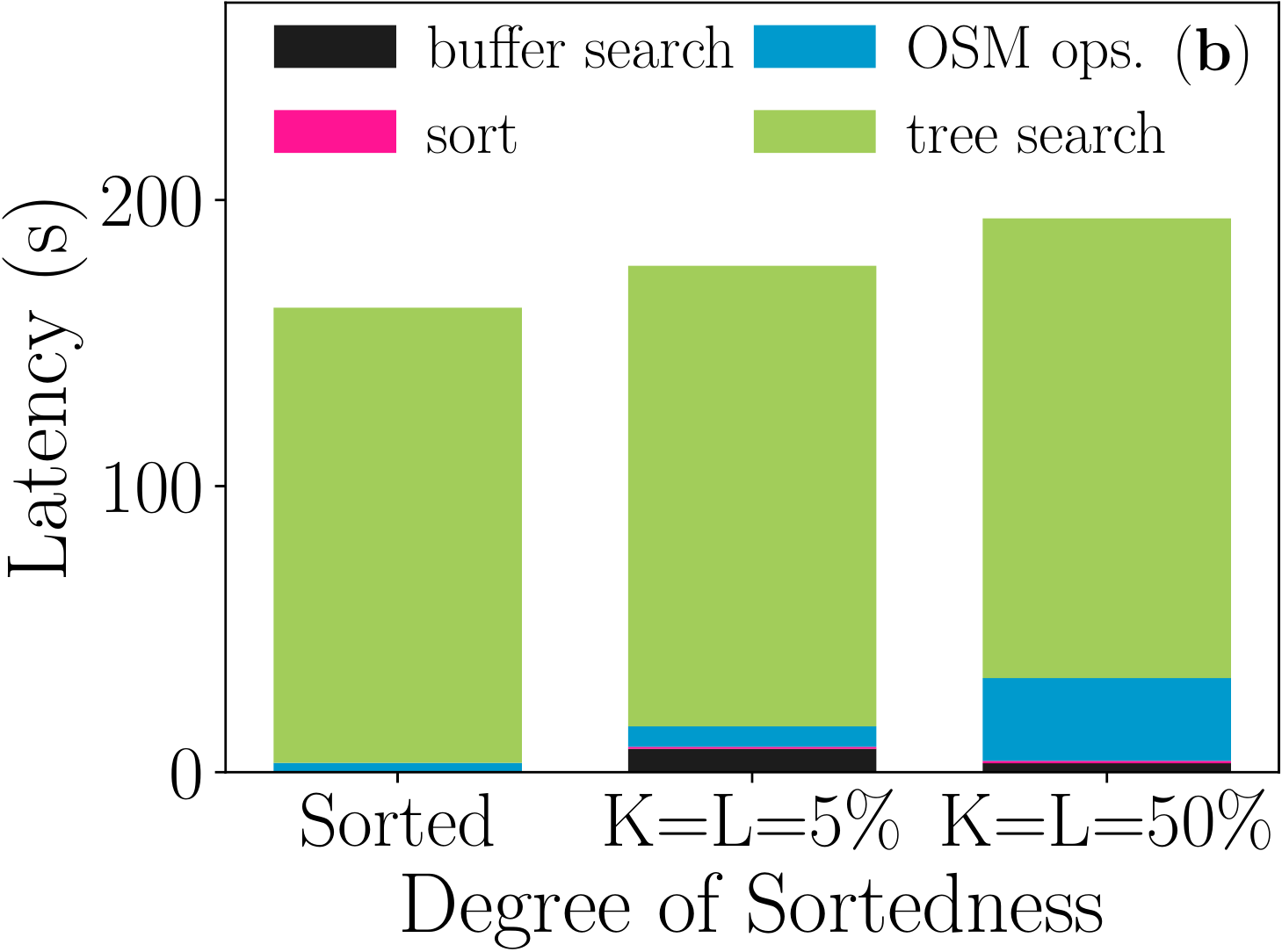}
		\label{}    
	\end{subfigure}    
	\vspace{-6mm}
	\caption{Latency Breakdown of Operations in \sysName{}. 
	(a) The time spent by \sysName{} for top-insert of entries into the index escalates with decreasing data sortedness. 
	(b) Tree search dominates query latency, while the fraction of time spent for metadata management and maintenance of zonemaps + BFs increases for lower data sortedness.}
    \label{fig:lat_breakdown}
    \vspace{-0.25in}
\end{figure}

\Paragraph{\sysName{} Dominates the Ingestion Performance}
Figure \ref{fig:raw_perf}(a) shows that \sysName{} performs significantly better than \bplustree{} for inserts if there is any degree of data sortedness. 
For fully sorted ($K$$=$$0\%$) and nearly sorted ($K$$=$$1\%$, $5\%$, and $10\%$) workloads, \sysName{} reduces the ingestion latency by $\sim$$90\%$ and $\sim$$82\%$, respectively. 
Even for data with lower degrees of sortedness ($K$$=$$25\%$ and $50\%$), the ingestion latency in \sysName{} is reduced by $\sim$$27\%$ compared to \bplustree{}. 
Figure \ref{fig:lat_breakdown}(a) shows the breakdown of the ingestion costs in \sysName{} for (i) fully sorted ($K$$=$$0\%$), (ii) nearly sorted ($K$$=$$L$$=$$5\%$), and less sorted ($K$$=$$L$$=$$50\%$) workloads. 
We observe that for fully sorted workloads, the \sysName{} is able to bulk load the entire data set without requiring any additional processing. 
For near-sorted data, \sysName{} sorts the buffer periodically (driven by the queries and the buffer-saturation) which accounts for $38\%$ of the workload execution latency. 
However, this additional cost paid to add structure to the data leads to significantly fewer top-inserts, which in turn, reduces the overall latency. 
Finally, for less sorted data, \sysName{} ingests fewer entries via bulk loading. Instead, a significant amount of data is ingested through top-inserts -- this behavior resembles the one of a \bplustree{}. However, as shown in 
Figure \ref{fig:mixed_perf}, \sysName{} still outperforms the state of the art by a significant margin.
Note that \sysName{} achieves this performance with a buffer 
that is 20\% in size when compared to $L$ (1\% vs. 5\%). 
This implies that even with a considerably small buffer that 
\emph{does not capture the out-of-order elements}, 
\sysName{} performs significantly fewer top-inserts and is able 
to bulk load a large fraction of the data.

\begin{table}[t]{
	\resizebox{0.475\textwidth}{!}{
	\begin{tabular}{l|cccl}\toprule 
	\multirow{2}{*}{Sortedness Degree} & \multicolumn{2}{c}{\# Nodes (\#Int + \#Leaf)} \\
		\cline{2-3}
		& \bplustree{} & \sysName{}  \\ \midrule
         Fully-Sorted & 2.004M (8K, 1.996M) & 0.52$\times$ (0.32$\times$, 0.52$\times$) \\
            Near-Sorted & 1.847M (7K, 1.840M) & 0.6$\times$ (0.38$\times$, 0.6$\times$) \\ 
            Less-Sorted & 1.878M (4.3K, 1.873M) & 1.01$\times$ (1.7$\times$, 1.01$\times$) \\  
		\bottomrule
    \end{tabular}
	}
}
    \caption{Memory footprint is reduces in \sysName{} with fully sorted and near-sorted data}
	\label{tab:memory_footprint}
     \vspace{-0.3in}
\end{table}

\Paragraph{Fast Ingestion Comes at a Small Overhead in Queries}
Figure~\ref{fig:raw_perf}(b) compares the point lookup performance of \sysName{} to that of the \bplustree{}, and we observe that \sysName{} incurs an overhead between $\sim$$5\%$ and $\sim$$26\%$ for point lookups. 
This is due to the additional time spent searching for the target entry in the buffer. 
Figure \ref{fig:lat_breakdown}(b) shows the breakdown of the point query latency in \sysName{} for (i) fully sorted ($K$$=$$0\%$), (ii) nearly sorted ($K$$=$$L$$=$$5\%$), and less sorted ($K$$=$$L$$=$$50\%$) workloads. 
We observe that regardless of the degree of data sortedness, between $80\%$ and $99\%$ of the query latency comes from the tree-search. 
With a full buffer, \sysName{} introduces an overhead due to (i) probing the \pinBufName{} using interpolation search and sequential scans and (ii) performing OSM-operations such as sort-merging the entries in buffer and updating the metadata. 
This overhead depends on the number of entries in the buffer and degree of data sortedness. 
Note that for fair comparison, we assumed the buffer to be completely full when the query workload was executed. 
In practice, the buffer is expected to be $50\%$ saturated on average, which would reduce the buffer-related overheads by $2\times$. 
Further, searching within the buffer and in the tree in parallel can reduce \sysName{}'s lookup cost. 

\Paragraph{The Benefits Outweigh the Overhead almost Always}
To weigh the benefits of \sysName{} against its read overhead, in Figure \ref{fig:raw_perf}(c), we show the mean latency per operation for a mixed workload. 
We observe that for a workload with equal number of reads and writes, \sysName{} improves the mean latency by $\sim$$70\%$ for fully and nearly sorted data. 
Even for workloads with larger degrees of sortedness, \sysName{} offers $1.25\times$ improved overall performance. 
To summarize, for read-only workloads, the performance of \sysName{} is similar to that of \bplustree{s}, as the buffer remains empty, and thus, adds no overhead. 
However, if a mixed workload is read-dominated (writes $<1\%$), the incurred read overhead out-weighs the benefits on ingestion of \sysName{}. 
We present a detailed account of the applicability of \sysName{} in Section \ref{subsec:workload}.


\Paragraph{OSM-Tree Offers Competitive Scan Performance}
Figure~\ref{fig:raw_perf}(d) shows that \sysName{} performs similarly to \bplustree{s} for range queries with different selectivity, varying from $0.01\%$ ($50$K entries) to $1\%$ ($5$M entries). We observe that scan latency for \sysName{} and \bplustree{} remain largely comparable while the selectivity is varied. 
This is because the scan latency is directly proportional to selectivity, and with a tree of the same height, the tree-traversal and leaf node scan costs are similar for both systems. 
For \sysName{}, the tree search latency dominates the scan time of the \pinBufName{}, which can be further reduced by parallelizing it with the tree search and sort-merging the result sets.

\vspace{-0.05in}
\subsection{Workload Influence} \label{subsec:workload}
\vspace{-0.05in}

\Paragraphit{Setup}
We measure the speedup offered by \sysName{} over state-of-the-art \bplustree{s} for mixed workloads 
with varying data sortedness. 
We vary both $K$$=$$(0, 1, 5, 10, 50)$ and $L$$=$$(1, 5, 10, 50)$ and experiment with $20$ different degrees of data sortedness. 

\begin{figure*}[t]
    \centering
    \hspace{-3mm}
    \begin{subfigure}[t]{0.23\textwidth}
        \centering
    	\includegraphics[scale=0.4]{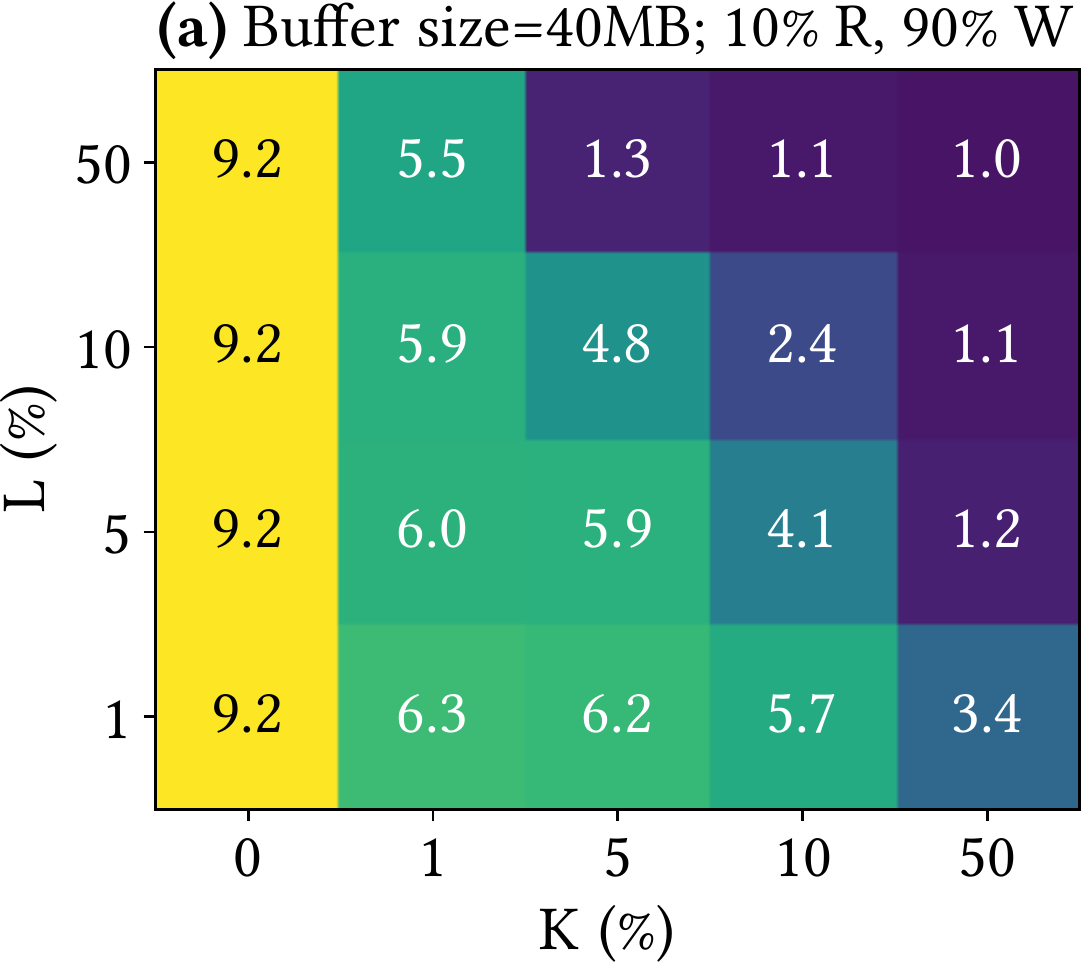}
    \end{subfigure}
    \hspace{4mm}
    \begin{subfigure}[t]{0.23\textwidth}
        \centering
    	\includegraphics[scale=0.4]{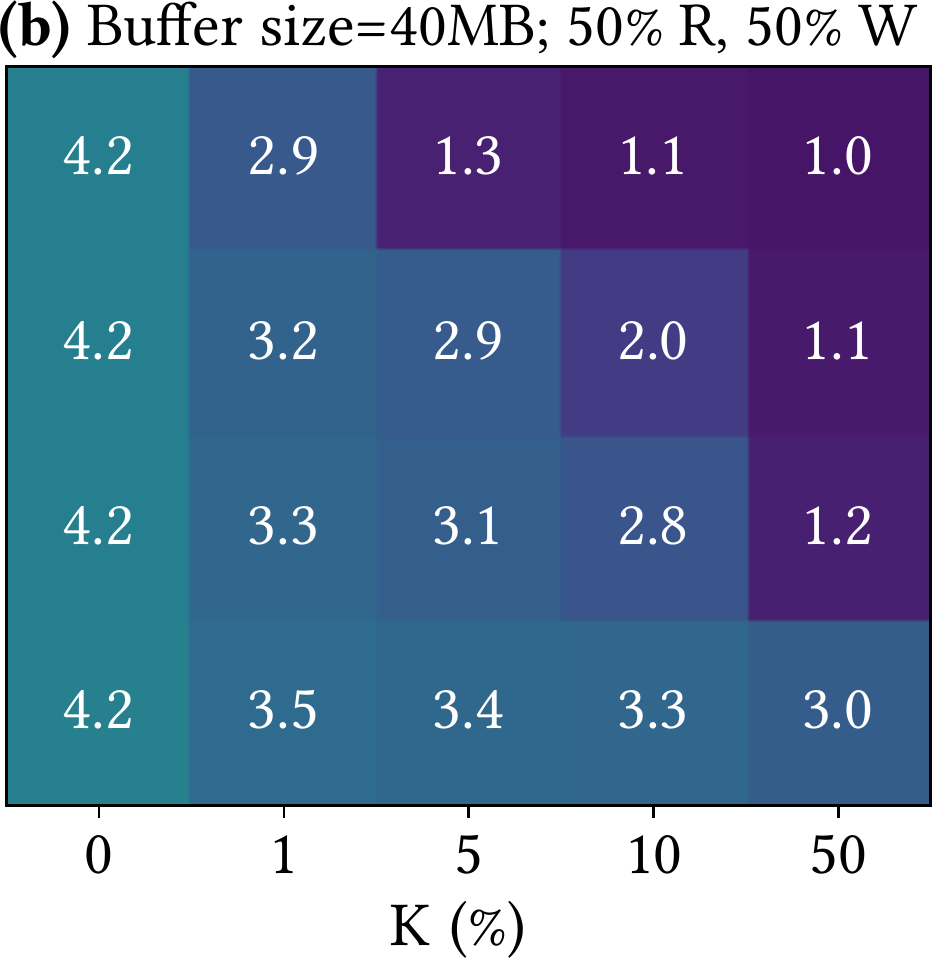}
    \end{subfigure}
    \hspace{-0.5mm}
	\begin{subfigure}[t]{0.23\textwidth}
        \centering
    	\includegraphics[scale=0.4]{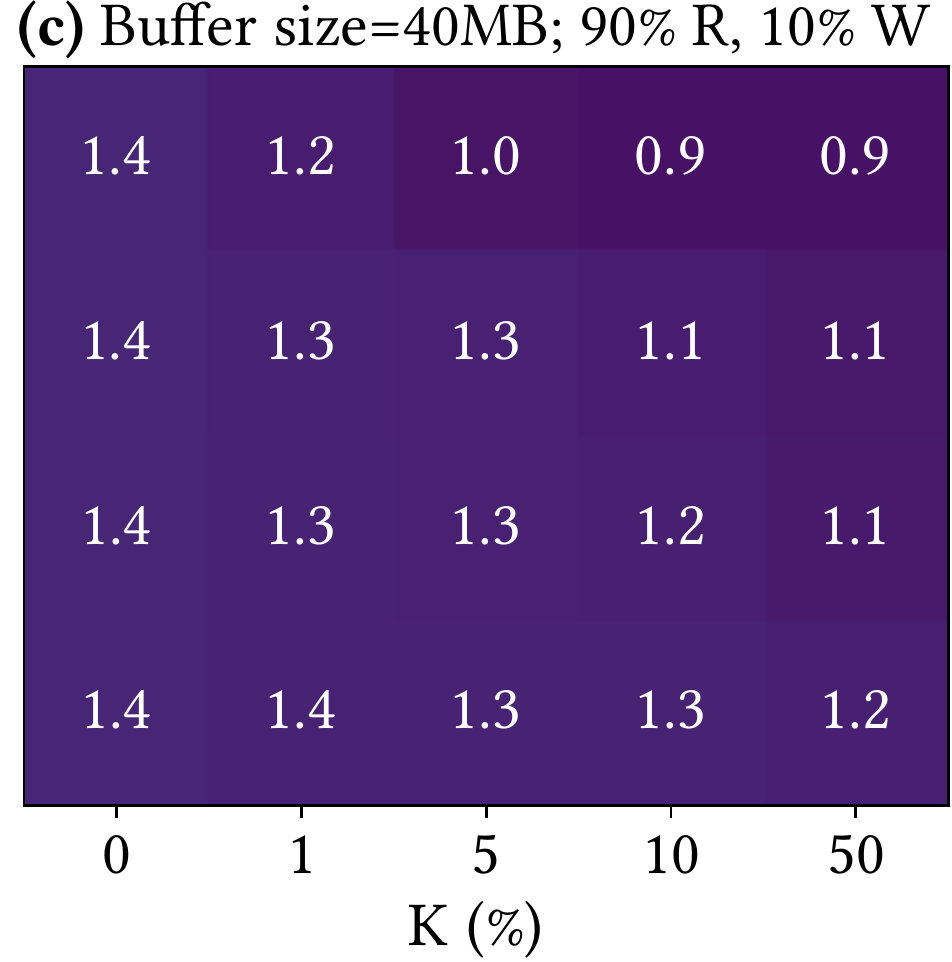}
    \end{subfigure}
    \hspace{1mm}
    \begin{subfigure}[t]{0.23\textwidth}
   \centering
    	\includegraphics[scale=0.4]{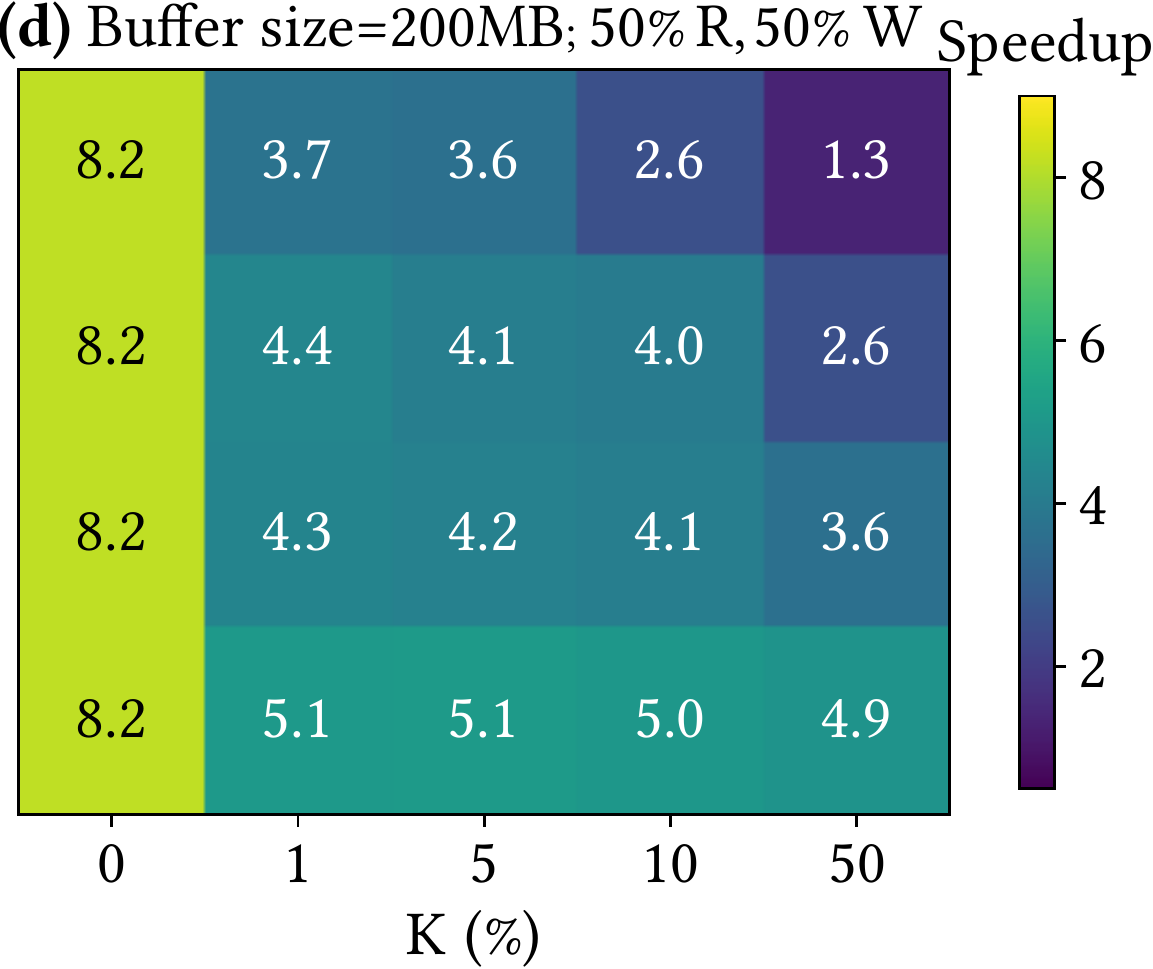}
	\end{subfigure}
    \vspace{-0.1in}
    \caption{Performance of \sysName{} with varying degrees of sortedness.
    (a) In a write-heavy workload, \sysName{} exploits data sortedness to offer maximum benefit in overal performance.
    (b) Increase in reads to the index diminishes the benefit offered by \sysName{}.
    (c) \sysName{} performs similar to \bplustree{} for read-heavy workloads with minimal performance benefits due to data sortedness. 
    (d) A larger buffer in \sysName{} is better at capturing even higher sortedness, to improve overall performance.}
    \label{fig:heatmaps}
\end{figure*}

\Paragraph{Varying the Workload Composition}
We observe in Figures \ref{fig:heatmaps}(a)-\ref{fig:heatmaps}(c) that as the proportion of reads increases in a workload (from $10\%$ to $50\%$ to $90\%$), the overhead incurred by reads in \sysName{} begins to counterbalance its ingestion-benefits. 
Even for a fully sorted workload, as the read-proportion increases from $10\%$ to $90\%$, the speedup is reduced from 9.2$\times$ to 1.4$\times$. 
Thus, for different degrees of data sortedness, Figure~\ref{fig:heatmaps} serve as a guideline for the applicability of the \sysName{} design. 

\Paragraph{Varying the Degree of Sortedness}
Analyzing the variation in the speedup for $K$$=$$1\%$ (second column) and $L$$=$$1\%$ (forth row) in Figure \ref{fig:heatmaps}(a), we observe that the effects on the number of unordered entries in the workload (i.e., $K$) influences the performance of \sysName{} by a greater extent compared to the maximum displacement of an unordered element (i.e., $L$). 
This is because if the buffer size is comparable to the $L$-value, the number of unordered entries drive the cost of data reordering, and relative overlaps between buffer cycles (causing top-inserts) is minimal.
However, as $L$ gets larger (first row), its impact on the speedup becomes more significant than that of $K$ (fifth column). 
With increase in both $K$ and $L$, the OSM-tree begins to operate similarly to \bplustree{s}, and the speedup offered approaches $1$. 


 \vspace{-0.05in}
\subsection{\pinBufName{} Tuning} \label{subsec:osm_tuning}
\vspace{-0.02in}

\Paragraphit{Setup} 
To analyze the implications of buffer size on \sysName{}'s performance, we first increase the buffer size to $5\%$ of the data size and compare the results with those discussed in Section~\ref{subsec:raw_perf}. 
Next, we run an ingestion-only workload with $500$M entries 
followed by a read-only workload with $50$M point lookups, and
we vary the buffer size between $0.5\%$-$5\%$ of the data size.  
For mixed workload, we pre-load the index (to $80\%$) and 
perform interleaved inserts and reads ($50\%W$-$50\%$R) for different degrees of sortedness.


\Paragraph{Increasing the Buffer Size Improves Performance} 
We now vary the size of the \pinBufName{} to show its implications on performance. 
The \pinBufName{} acts as a reservoir that absorbs the out-of-order ingested elements to an extent. 
Thus, increasing the buffer size allows for bulk loading a larger fraction on the data while reducing the number of top-inserts. 
Figure~\ref{fig:heatmaps}(d) shows the speedup offered by \sysName{} when \pinBufName{} size is increased to $5\%$ of the data size ($200$MB) for a mixed workload with equal proportions of reads and writes. 
Comparing with Figure~\ref{fig:heatmaps}(b) (which has the same setup but a buffer size of $1\%$), we observe that a $5\times$ increase in the buffer size increases the speedup  from $4.2\times$ to $8.2\times$ (a $95.2\%$ increase) for a fully sorted workload and between $27.6\%$ and $176.9\%$ for nearly sorted data. 
Even for $K$$=$$L$$=$$50\%$, increasing the buffer size improves the achieves speedup from $1\times$ to $1.3\times$.

\begin{wrapfigure}{r}{0.24\textwidth}
	\vspace{-0.05in}
	\centering
	\hspace{-0.15in}
    \includegraphics[scale=0.32]{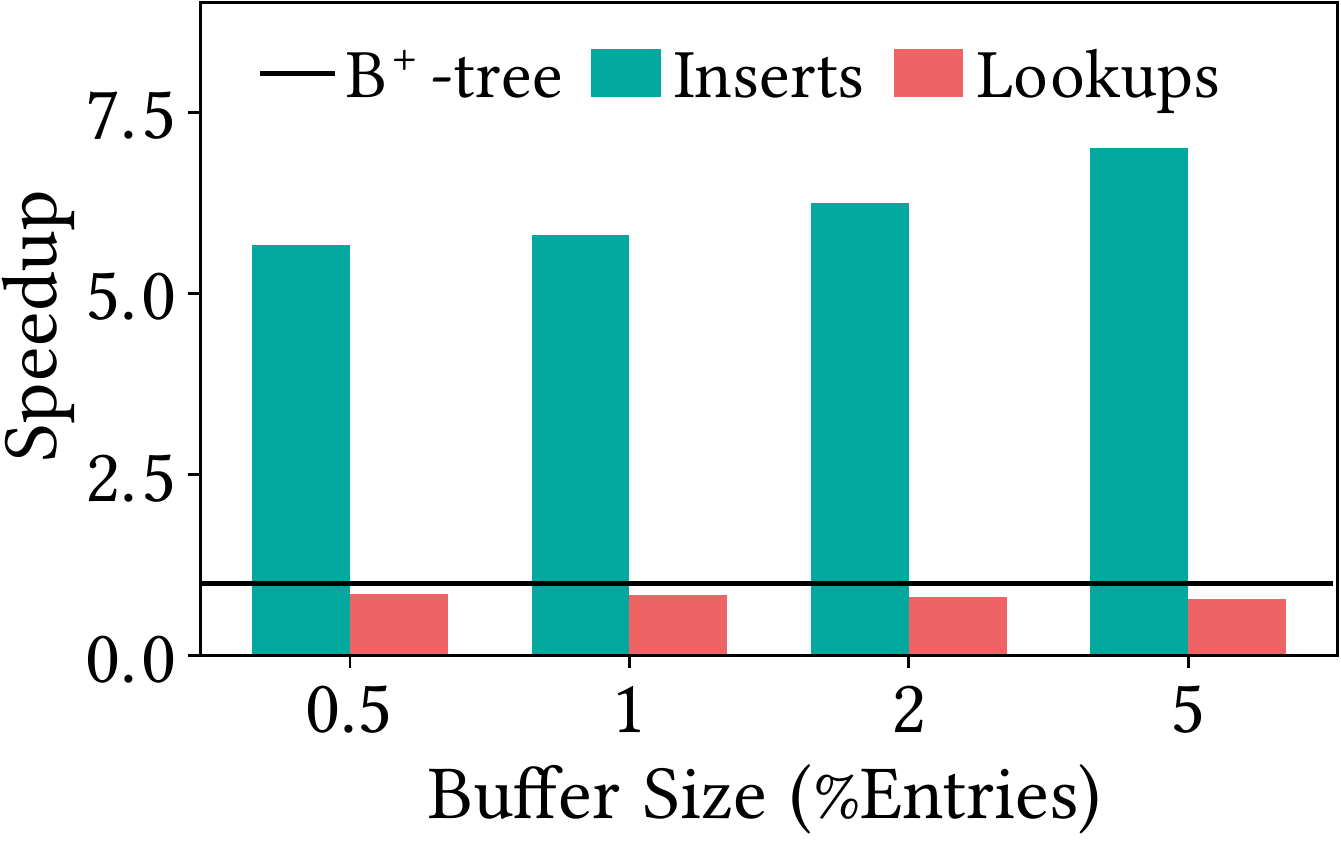}
	\vspace{-0.1in}
    \caption{The ingestion performance of \sysName{} increases proportionally with buffer size.}
    \label{fig:osm_vary_buffer}
	\vspace{-0.15in}
\end{wrapfigure}

\Paragraph{Buffer Size Affects the Ingestion Performance Significantly}
Figure~\ref{fig:osm_vary_buffer} shows the implications of varying buffer size separately on the ingestion and lookup performance for \sysName{}. 
In this set of experiments, we vary the \pinBufName{} size for a fixed sortedness ($K=L=5\%$).
We observe that even with a small buffer size of $20$MB ($0.5\%$ of the data size), \sysName{} offers a $5.7\times$ speedup during ingestion. 
As the buffer size increases to $5\%$, the ingestion speedup further increases to $7\times$, since a larger buffer reduces the proportion of data ingested through top-inserts. 
\setlength{\columnsep}{12pt}
\setlength{\intextsep}{8pt}
We also observe in Figure ~\ref{fig:osm_vary_buffer} 
that point lookup performance in \sysName{} is affected by the buffer size marginally. 
A $10\times$ increase in the buffer size (from $20$MB to $200$MB) induces a deterioration of $11.5\%$ in the lookup latency. 
This conforms with our observations from Figures~\ref{fig:raw_perf}(b) and~\ref{fig:heatmaps}(c) regarding the suitability of the OSM-design;
however, as shown in Figure~\ref{fig:mixed_perf} the benefits on \sysName{} outweigh the read-overheads even for a small 
fraction of writes ($\geq$$5\%$).

\Paragraph{Tuning the Buffer Flush Threshold}
Adjusting the flush threshold of the \pinBufName{} affects the overall performance of \sysName{}. 
We now vary the proportion of entries flushed from the buffer at a given cycle between $25\%, 50\%$, and $75\%$, and run mixed workloads. 
In the interest of space, we omit the figure and focus on the key observation.
When the buffer flush threshold is set to $25\%$, \sysName{} offers a speedup between $1.0\times$ and $4.0\times$. 
For flush threshold $50\%$, the speedup of \sysName{} becomes between $1.0\times$ and $4.3\times$, and for a threshold of $75\%$, between $0.91\times$ and $4.2\times$. 
Hence, \sysName{} performs best for 
$50\%$ flush threshold, which we default to.

\begin{wrapfigure}{r}{0.24\textwidth}
	\vspace{-0.05in}
	\centering
	\hspace{-0.15in}
    \includegraphics[scale=0.32]{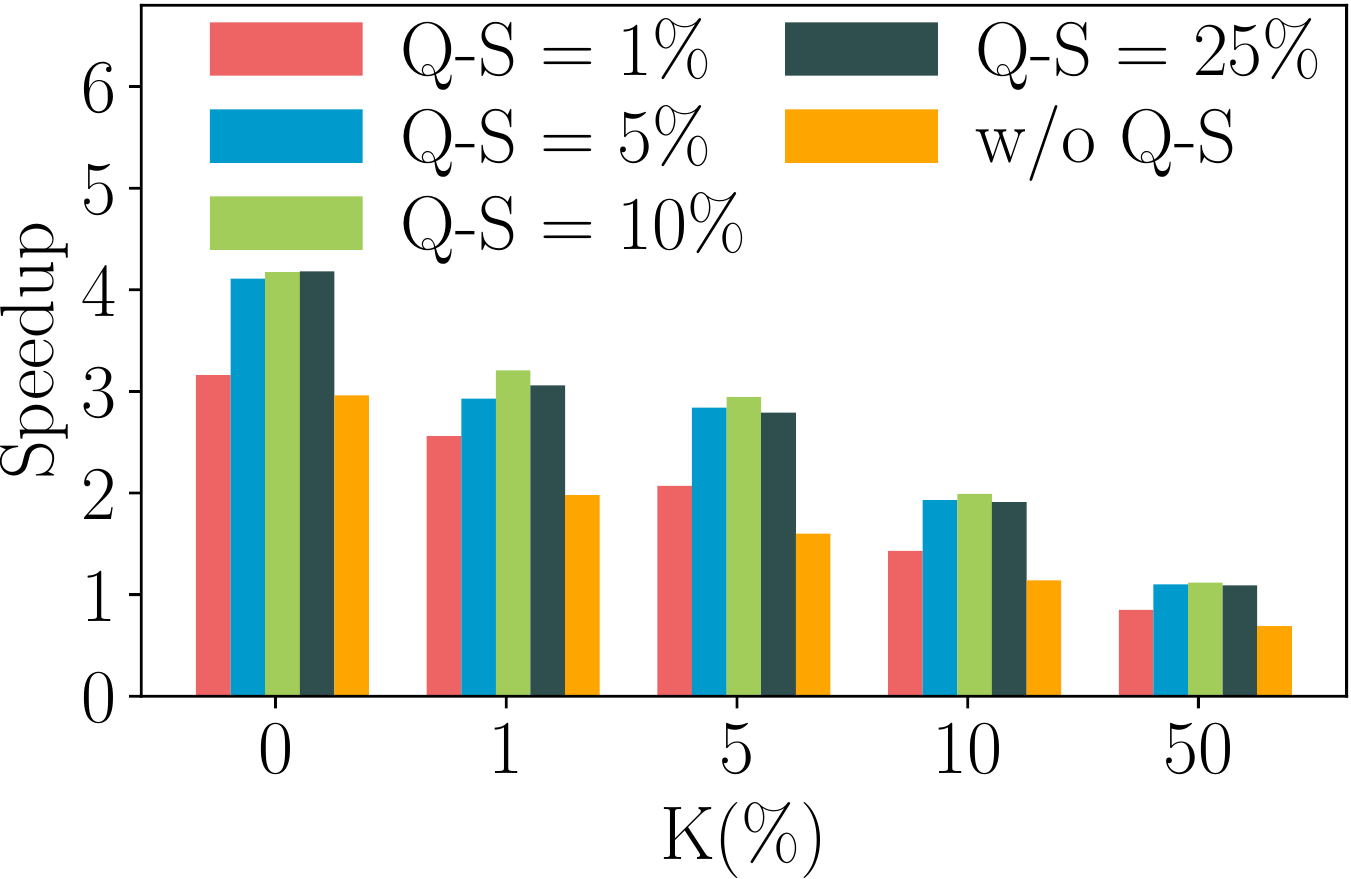}
	\vspace{-0.15in}
    \caption{Query-based sorting threshold set to $10\%$ offers the highest speedup.}
    \label{fig:osm_qs_configs}
	\vspace{-0.1in}
\end{wrapfigure}

\Paragraph{Tuning Query-Based Sorting}
Figure \ref{fig:osm_qs_configs} shows the implications of query-based sorting on the overall performance of \sysName{}. 
We
\setlength{\columnsep}{10pt}
\setlength{\intextsep}{5pt}
vary the query-based sorting threshold between $1\%$, $5\%$, $10\%$, $25\%$ 
and $100\%$ (disabling query-based sorting) and run mixed workloads. 
The y-axis
shows the speedup when compared against our \bplustree{} baseline.
We observe that 
enabling query-based sorting offers a performance improvement between 
$7\%$ (for $1\%$ threshold) and $25\%$ (for $10\%$ threshold). As expected by
gradually sorting the buffered data significantly accelerates query performance
on average since the 
portion the buffer that needs to be scanned is kept small.
Moreover, we observe that query-based sorting has diminishing returns if we
apply it too frequently.
Specifically, 
setting the threshold to $10\%$ offers the maximum speedup for any data sortedness, while other values affect 
performance adversely (if at all). Reducing the threshold (to $1\%$ or $5\%$) leads to too much sorting, while increasing the threshold (to $25\%$) results in fewer sorted blocks within the buffer, 
so the cost of scanning the unsorted section remains high. Hence, we emperically tune \sysName{} to perform query-based sorting with threshold $10\%$.

\begin{figure}[t]
    \centering
	\hspace{1mm}
    \begin{subfigure}[t]{0.2\textwidth}
        \centering
        \hspace{-0.4in}
		\includegraphics[scale=0.28]{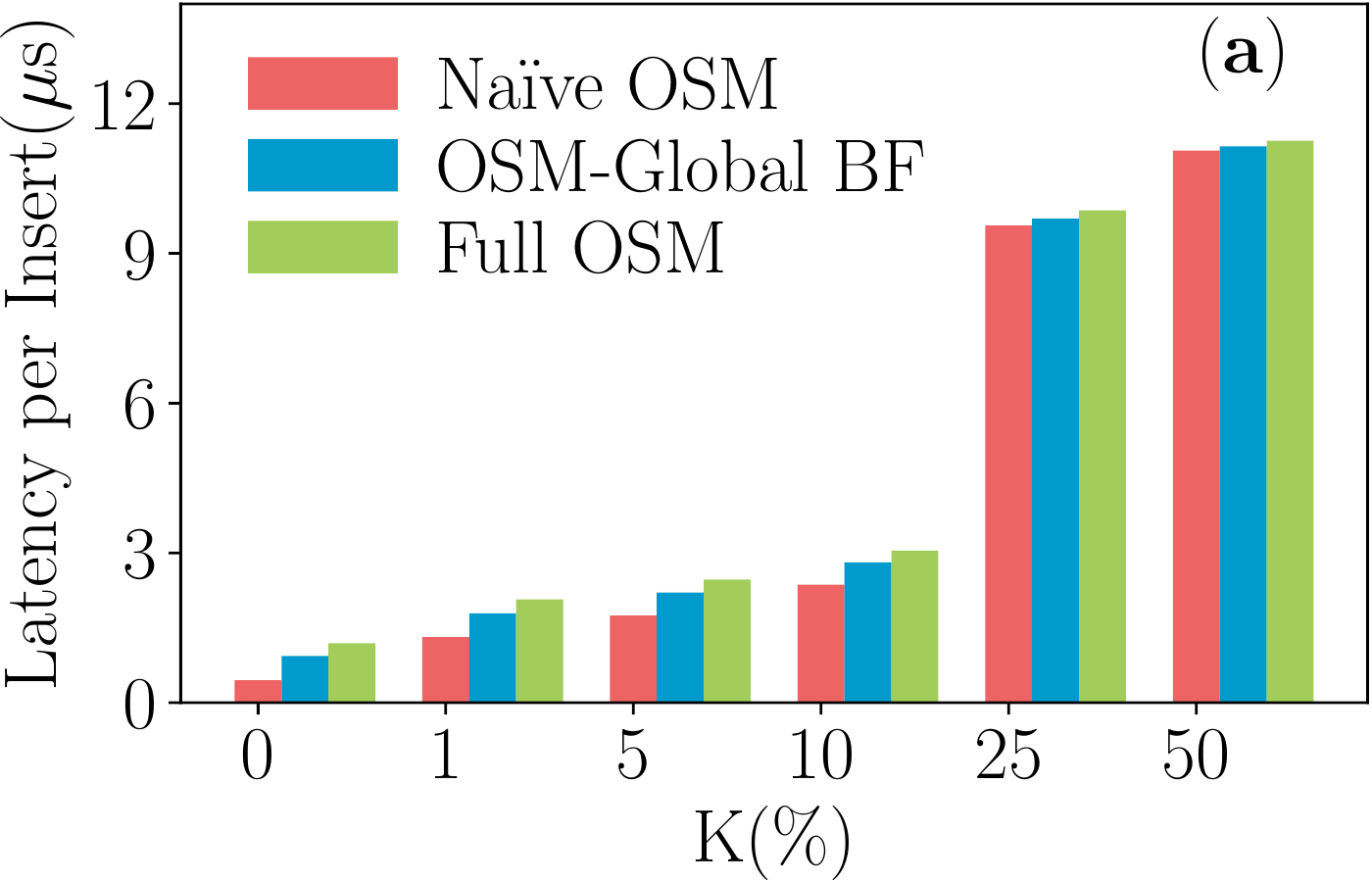}
		\label{}
    \end{subfigure}
	\hspace{-3mm}
    \begin{subfigure}[t]{0.2\textwidth}
      \centering
		\includegraphics[scale=0.28]{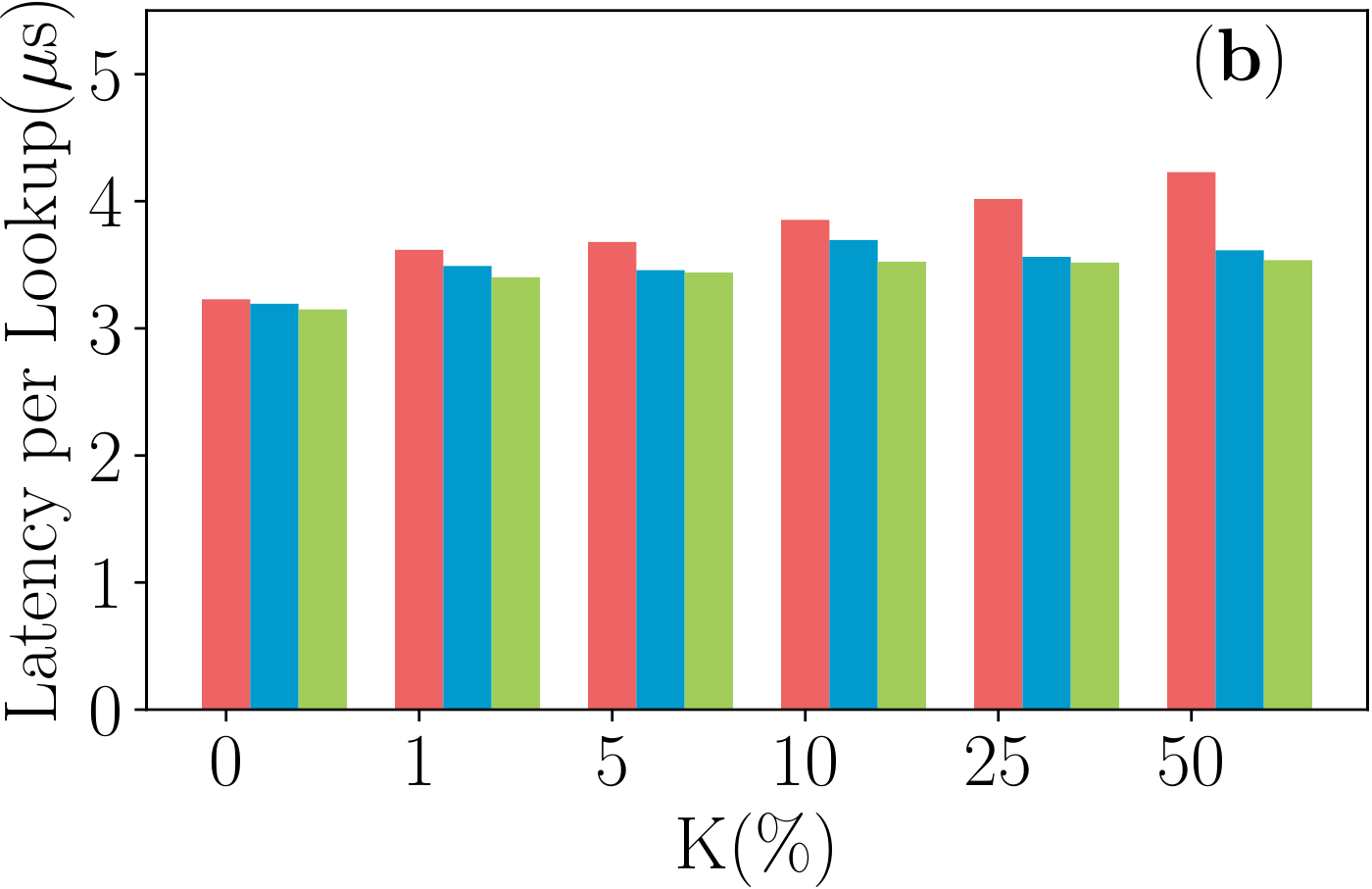}
		\label{}    
	\end{subfigure}    
	\vspace{-0.25in}
	\caption{Latency breakdown for different OSM-tree configurations. 
	(a) Adding BFs to the \pinBufName{} slightly increases the insert latency. 
	(b) The use of BFs in the buffer for lookups is more pronounced as data sortedness decreases.}
    \label{fig:bfs_variations}
	\vspace{-0.15in}
\end{figure}

\Paragraph{Benefits from the Global and Per-page BFs}
To demonstrate the benefits coming from the global BF and from the per-page
BFs in the \pinBufName{} we compare \sysName{} with two simpler variations:
one without any BFs (\emph{Na\"ive OSM}) and one with only the global BF
(\emph{OSM-Global BF}) by ingesting $500$M near-sorted entries and then 
performing $50$M non-empty point lookups. 
As expected, updating the BFs during every insertion comes at a marginal cost 
increase at ingestion time as shown 
in Figure \ref{fig:bfs_variations}(a). For less sorted data the added cost at 
ingestion time is a smaller fraction of the total insert time.
However, the additional cost at ingestion time pays off since at query time we
have significant performance benefits. 
Figure \ref{fig:bfs_variations}(b) shows that adding the global BF speeds up
queries up to $14\%$, while the per-page BFs further boost performance leading
to $16\%$ aggregate improvement. The positive impact of per-page BFs
is limited by query-based sorting that also helps avoid scanning unnecessary
data (limiting the portion of \pinBufName{} to be scanned to $<$$10\%$).
%
%
Note that the ingestion performance of 
\sysName{} is always significantly faster than 
\bplustree{} regardless of the BF cost. 
Overall, the benefit of BFs is pronounced for workloads with more
reads.



\Paragraph{Tuning Zonemaps}
\pinBufName{} uses Zonemaps during ingestion to approximate sortedness 
(\S\ref{sec:w-opt-buff-tree}) making them integral to the overall design. While
we opt to always use them at query time since they are always available, we experimentally measured that skipping
Zonemaps at query time reduces performance by 35\% 
on average.

\subsection{On-Disk Performance}
\begin{figure}[t]
    \centering
    \includegraphics[scale=0.45]{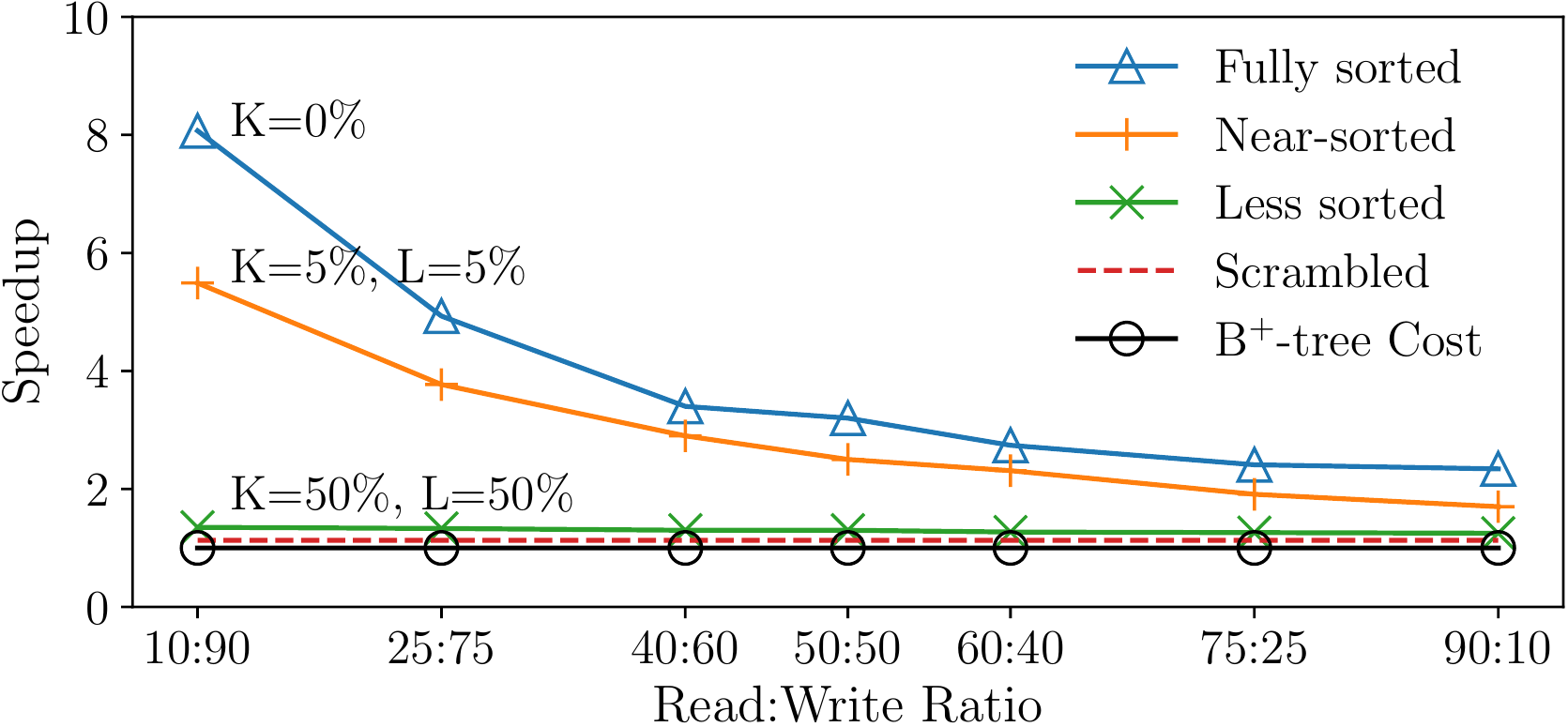}
    \vspace{-3mm}
    \caption{For disk-based execution with 1\% bufferpool, \sysName{} always outperforms \bplustree{}.}
    \label{fig:mixed-perf-disk}
    \vspace{-3mm}
\end{figure}

In our next experiment, we explore an \sysName{} setup that accesses disk-resident
data. Our design comes with a bufferpool that, in this experiment, fits all internal tree nodes ($\sim$$1\%$ of data size).
We repeat the initial experiment for variable data sortedness
and variable read vs. write ratio, to present the speedup between \sysName{}
and \bplustree{} as shown in Figure \ref{fig:mixed-perf-disk}. Note that the \pinBufName{} is set to 
$1\%$ of the total data. From the disk-based experiment we draw similar conclusions
to the initial in-memory one (Fig.~\ref{fig:mixed_perf}), with a notable 
difference. \sysName{} now \emph{always} outperforms
\bplustree{} even for read-intensive workloads with fully scrambled data.
This is because, regardless of data sortedness, we increase locality through our 
sorting procedures in the buffer.
Though this is applicable for both in-memory and disk-based experiments with \sysName{}, 
the overhead of managing the buffer is negligible 
when compared to accessing tree nodes for the disk-based experiment. 
Overall, when spilling to disk, \sysName{} leads to significant performance savings
up to $8\times$ for write-intensive workloads with high data sortedness, while
always outperforming \bplustree{}.

\subsection{Scalability}
\begin{figure}[t]
    \centering
    \begin{subfigure}[b]{0.24\textwidth}
        \centering
    \includegraphics[scale=0.3]{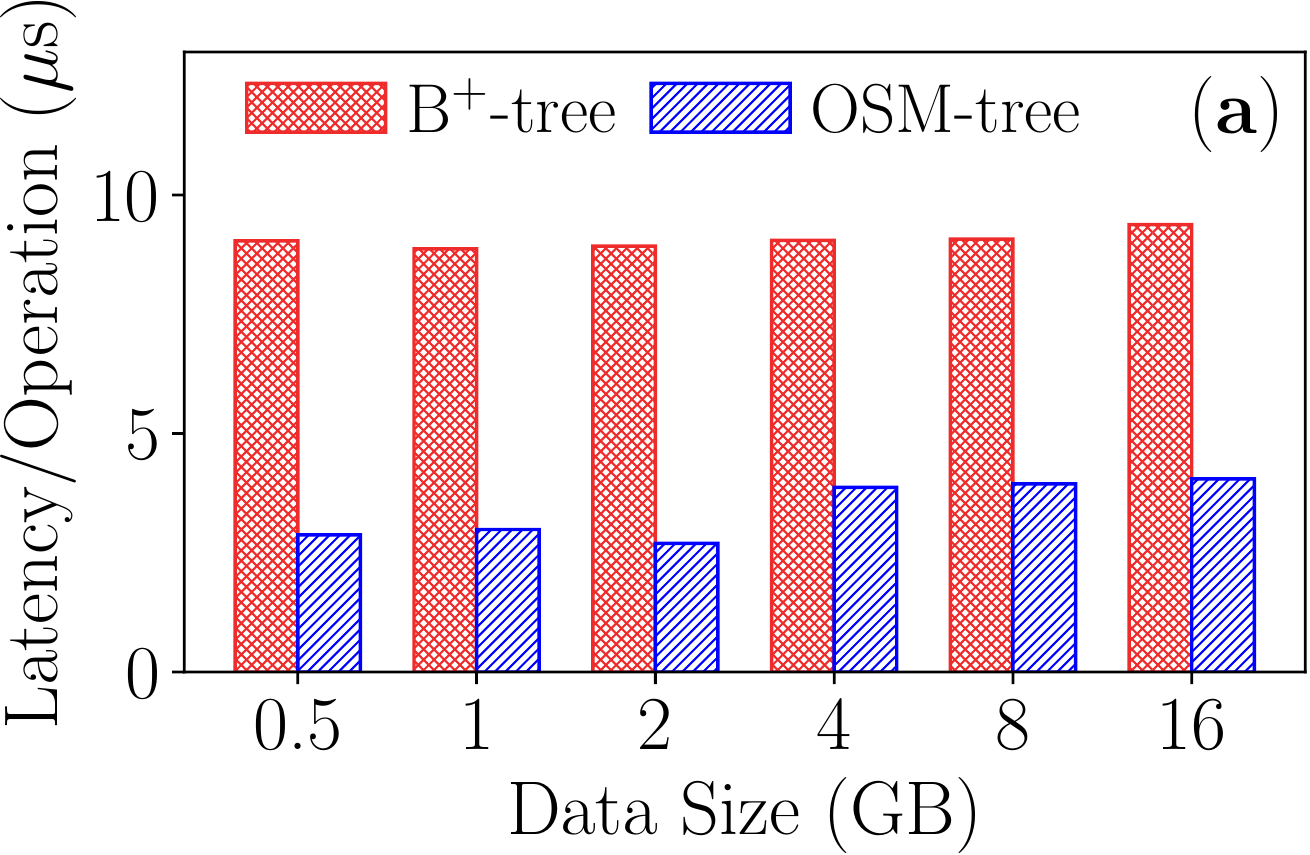}
    \end{subfigure}
    \hspace{-3mm}
    \begin{subfigure}[b]{0.24\textwidth}
        \centering
    \includegraphics[scale=0.3]{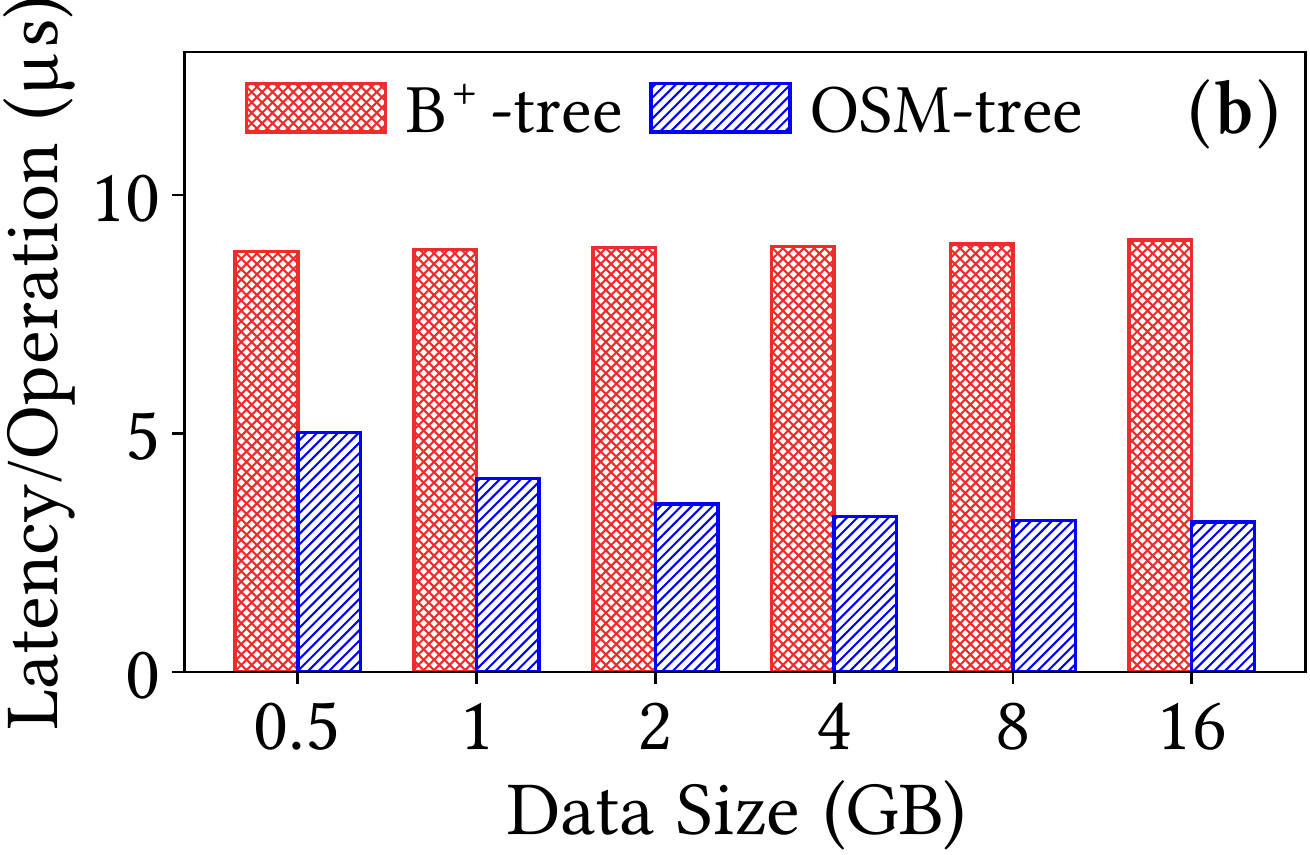}
    \end{subfigure}
	\vspace{-3mm}
    \caption{\sysName{} scales well with data size and always outperforms \bplustree{}.}
\label{fig:scalability}
\vspace{-0.05in}
\end{figure}

\Paragraph{OSM-Tree Scales Similarly to the State of the Art}
To analyze the scalability of \sysName{}, we increase the number of entries ingested from $31.25$M to $1$B while varying $K$ and $L$ as proportional ($5\%$) to the workload size. 
We also scale the actual size of \pinBufName{} by keeping it equal to $1\%$ of the dataset size. 
For each experiment, we first ingest $80\%$ of the ingestion workload and then interleave the remain inserts with equal number of of point lookups. 
As expected, the degree of data sortedness does not affect the performance of \bplustree{s} and the latency per operation remains nearly unaffected by the increase in data size. 
\sysName{} scales similarly to the state of the art while offering a speedup between $2.32\times$ and $3.14\times$ as shown in Figure~\ref{fig:scalability}(a). 
Increasing the buffer size proportionally with $L$ allows \sysName{} to absorb the same degree of sortedness within the buffer regardless of the buffer size. 

\Paragraph{OSM-Tree Scales Better for Fixed $L$ and Buffer Size}
In this experiment, we treat $L$ as a constant rather than as a fraction of the data set. We set $L$ to $12.5$M and the buffer size 
to $40$MB. We then run a mixed workload with equal reads and writes on a preloaded index. In 
Figure~\ref{fig:scalability}(b), we observe that while \bplustree{} scales steadily with the data size (as explained earlier), as we increase the number of entries, the mean latency per operation reduces for \sysName{}. 
\sysName{} offers between 43\% and 65\% improvement in terms
of the latency per operation compared to \bplustree{}. 
A $32\times$ increase in the data size reduces the average 
query latency by $\sim$$37\%$. 
As the data size increases, the 
fraction of the ingested data retained in 
the buffer reduces exponentially (as shown in Table \ref{tab:scalability_fixed_l_and_buff}). 
This causes an increasing fraction of point lookups to skip the 
buffer through the \pinBufName{} Zonemap, hence, avoiding the
\pinBufName{} access cost.

\subsection{Experimenting with TPC-H}
Finally, we evaluate \sysName{} by comparing its performance against state-of-the-art \bplustree{s} on TPC-H~\cite{TPCH} data. 

\Paragraphit{Setup}
For this experiment, we use the tuples in the \texttt{lineitem} table (as discussed in Section~\ref{sec:introduction}) as our workload. 
We sort the tuples based on the \texttt{shipdate} attribute 
which, in turn, creates a nearly sorted data set with respect to the \texttt{receiptdate} attribute. 
We attribute this degree of sortedness on \texttt{receiptdate} as: $K$$=$$96.67\%$ and $L$$=$$0.1\%$.
The workload is comprised  of $6$M tuples. 
For both \sysName{} and the \bplustree, we first preload the database and the index with $4.8$M entries and then execute a mixed workload with varying proportions of reads and writes. 
We measure the latency per operation during the mixed workload execution and compute the speedup accordingly. 
For each set of experiment, we also vary the buffer size between $0.05\%$ and $1\%$ of the data size.

\begin{table}[t]
	\resizebox{0.475\textwidth}{!}{
	\begin{tabular}{l|ccccccl}\toprule 
		Datasize & $0.5$GB & $1$GB & $2$GB & $4$GB & $8$GB & $16$GB \\ \midrule
		\#Entries & $31.25$M & $62.5$M  & $125$M  & $250$M & $500$M & $1$B \\
		\#Ent. in Buff. & $2.5$M & $2.5$M  & $2.5$M  & $2.5$M & $2.5$M & $2.5$M \\
		\%Ent. in Buff. & $8\%$ & $4\%$  & $2\%$  & $1\%$ & $0.5\%$ & $0.25\%$ \\
		\#Ent. in Tree & $28.75$M & $60$M  & $122.5$M  & $247.5$M & $497.5$M & $997.5$M \\
		\bottomrule
    \end{tabular}
	}
    \caption{For a fixed $L$ and buffer size, the fraction of entries in the buffer reduce with increasing workload data size.}
	\label{tab:scalability_fixed_l_and_buff}
	\vspace{-0.35in}
\end{table}

\Paragraph{OSM-Tree Offers Superior Performance} 
Table~\ref{tab:tpch} shows that \sysName{} performs significantly better than \bplustree{s} across all buffer sizes and for all workload compositions. 
Even with a buffer that is $0.05\%$ of the data size, \sysName{} offers between $1.14\times$ and $1.63\times$ speedup. 
As the buffer size increases, it is able to cache more entries before flushing, which reduces the number of top-inserts performed, improving ingestion performance. 
We also observe that the benefits of \sysName{} diminish as the proportion of reads increase in the workload; however, even for a workload with $90\%$ reads, \sysName{} offers a speedup of $1.3\times$ on average. 
Interestingly, for larger proportions of reads ($\geq$$75\%$), 
a larger buffer size ($\geq$$1\%$) causes reads to probe more 
data in the buffer for every lookup, which, in turn, causes a 
slight drop in \sysName{}'s speedup.  
Overall, this experiment highlights that the OSM-design is able 
to offer significant performance benefits compared to the state 
of the art with a very small buffer size ($0.05\%$) even for 
workloads with large proportions of reads ($90\%$) and low 
degree of data sortedness ($K$$=$$96.67\%$).

	\begin{table}[t]
			\resizebox{0.45\textwidth}{!}
			{
			\begin{tabular}{lccccccl}\toprule 
			\multirow{2}{*}{Read : Writes} & \multicolumn{5}{c}{Buffer Size (\%data size )}\\ 
			\cline{2-6}
			  & $0.05\%$ & $0.1\%$ & $0.25\%$ & $0.5\%$ & $1.0\%$ &	 \\
			\midrule
			$10\%$ : $90\%$ & $1.63\times$ & $2.60\times$ & $2.88\times$ & $4.46\times$ & $5.28\times$\\
			$25\%$ : $75\%$ & $1.54\times$ & $2.40\times$ & $2.49\times$ & $3.62\times$ & $4.57\times$\\
			$50\%$ : $50\%$ & $1.56\times$ & $2.07\times$ & $2.82\times$ & $3.21\times$ & $3.40\times$ \\
			$75\%$ : $25\%$ & $1.25\times$ & $1.65\times$ & $1.72\times$ & $2.09\times$ & $2.01\times$ \\
			$90\%$ : $10\%$ & $1.14\times$ & $1.24\times$ & $1.28\times$ & $1.42\times$ & $1.41\times$ \\
			\bottomrule
		   \end{tabular}
			}
		   \caption{When querying TPC-H data, \sysName{} always outperforms \bplustree{} offering speedups between $1.14\times$ and $5.3\times$.}
		\label{tab:tpch}
		\vspace{-0.1in}
	\end{table}

\section{Related Work}
\label{sec:related}

While to the best of our knowledge, this is the first work on designing sortedness-aware indexes, in this section, we discuss the literature on ingestion-optimized index structures.

\Paragraph{Optimizing for Tree Ingestion}
\bplustree{s} are widely used as the indexing data structure in commercial database systems due to its balanced ingestion and query performance~\cite{Comer1979}. 
Over the past years, several \bplustree{}-variants have been proposed that focus on optimizing ingestion and promote batching. 
Lehman and Carey proposed an in-memory tree index, T-tree~\cite{Lehman1986}, which improves insertion and access performance by storing pointers to data in the nodes. 
$\text{CSB}^+$-tree~\cite{Rao2000} and PLI-tree~\cite{Torp1998} attempt to maximize the cache line utilization by using arithmetic operations to calculate the child nodes rather than pointer operations to reduce the ingestion latency. 
YATS-tree (or Y-tree)~\cite{Jermaine1999} is a hierarchical indexing structure that aims to maximize bulk insertion by pushing new inserts into separate blocks based on a total order. 
Partitioned \bplustree{}~\cite{Graefe2003} 
optimizes bulk insertion by using an artificial leading 
column to always append, which leads to creating multiple 
indexes on overlapping data.
At query time, in case the leading column (e.g., epoch) is 
known, this leads to efficient execution, otherwise, 
multiple partitions have to be searched. 
Similarly, \bepsilontree~\cite{Brodal2003} reduces the cost of trickle insertions by buffering data in internal nodes.

Contrary to \sysName{}, none of the above \bplustree{}-variants aim to exploit implicit structure in the ingestion workload. 
The \sysName{} design techniques presented in this paper, 
offer superior performance by allowing efficient out-of-order insertions and taking advantage of data sortedness when available. 
Furthermore, the OSM design paradigm can be adapted to reduce storage overhead and cache misses, as well as to reduce the balancing cost.

\balance 
\Paragraph{LSM-trees}
LSM-trees~\cite{ONeil1996,Luo2018} optimize data ingestion by 
buffering and flushing to disk sorted pages, 
however, their design comes at a high write amplification cost. 
Data entries are repeatedly re-written on disk, and
LSM-trees periodically sort-merge smaller sorted components to create 
larger sorted collections of data through the process of 
\textit{compactions} \cite{Sarkar2021c}.
While LSM-trees aim to maximize ingestion throughput, they are not
designed to exploit sortedness. In fact, most LSM-tree designs are
completely agnostic to data sortedness and they perform the same amount
of merging and (re-)writing of the data on disk even when the data arrive
fully sorted. When an LSM-tree employs a leveled partial compaction 
strategy with \emph{no-overlap} data movement policy \cite{Sarkar2021c} it 
can accelerate ingestion of sorted data, however, to fully exploit 
a varying degree of sortedness, more changes are needed in various
compaction decisions. LSM can benefit from the OSM meta-design to better
exploit variable sortedness. Further, the LSM design \emph{per se} can
be optimized to better handle sorted data ingestion.


\Paragraph{Data Series and Data Streaming}
Data series store data with a monotonically 
increasing component, typically a timestamp \cite{Palpanas2015}. Data
series indexing assumes that data ingestion 
follows the expected 
order~\cite{Kondylakis2018,Zoumpatianos2014,Zoumpatianos2016,Zoumpatianos2018}.
The ingested data is converted to shapes using specialized representations 
like iSAX~\cite{Camerra2010}, in order to allow similarity comparisons 
between data series. 

Data streaming applications operate on windows of data (typically time-based)
in order to calculate state on-the-fly, and then, discard the incoming 
entries~\cite{Carbone2020,Fragkoulis2020}. Hence, streaming systems 
inspect whether data arrives out of the expected order and often use a buffer to 
capture this arrival skew \cite{Srivastava2004}. They do not build an index for the entire dataset, and similar to data
series, the default expectation is that data arrives in the expected order. 

Contrary to data series and data streaming, in relational systems, the
arrival of data is, in general, scrambled; 
however, indexes are not designed to benefit from data arriving in-order or near-order.
In this work, we \emph{treat 
sortedness as a resource}, and we build a general purpose
index that can substantially outperform existing indexes if data ingestion 
order follows the indexed key, while falling back to baseline 
performance if there is no underline data sortedness.

\balance
\section{Conclusion}
\label{sec:conclusion}
Inserting data to an index can be perceived as
the process of adding structure to an otherwise 
unsorted data collection. We identify 
inherent data sortedness as a resource that 
should be harnessed when ingesting data. 
However, state-of-the-art index designs like \bplustree{s} can only enable faster 
ingestion through bulk loading when data arrives fully sorted, and they fail to benefit 
from sortedness when data is near-sorted. 

To address this, we propose an index meta-design
that allows for progressively faster ingestion 
when the incoming data has increasingly higher 
sortedness. Our proposed design, called 
\sysName{}, uses partial bulk loading, index appends, 
variable split factor, and in-memory buffering to 
amortize the insertion cost. \sysName{} also 
offers competitive lookup performance by 
engaging Bloom filters and Zonemaps for data in the 
in-memory buffer. Our experiments demonstrate that 
\sysName{} outperforms the baseline \bplustree{} 
by up to $8.8\times$ in presence of data sortedness 
and offers up to $5\times$ 
performance improvement for mixed read/write 
workloads. 

\balance
\newpage
\bibliographystyle{abbrv}
{
\bibliography{../../../Dropbox/W-Lab/Bibliography-Mendeley/library, bibl}
}

\end{document}